\documentclass[12pt]{article}

\usepackage{epsf}
\usepackage{latexsym,amssymb,euscript}
\usepackage[dvips]{graphicx}
\usepackage{amsmath}

\begin{document}

\begin{center}
{\Large \textbf{Numerical study of the phase transitions in the 
two-dimensional $Z(5)$ vector model}}
\end{center}

\vskip 0.3cm
\centerline{O. Borisenko$^*$}
\vskip 0.3cm
\centerline{\sl Bogolyubov Institute for Theoretical Physics,}
\centerline{\sl National Academy of Sciences of Ukraine,}
\centerline{\sl 03680 Kiev, Ukraine}
\vskip 0.3cm
\centerline{G. Cortese$^{**}$, R. Fiore$^{**}$}
\vskip 0.3cm
\centerline{\sl Dipartimento di Fisica, Universit\`a della Calabria,}
\centerline{\sl and Istituto Nazionale di Fisica Nucleare, Gruppo collegato
di Cosenza}
\centerline{\sl I-87036 Arcavacata di Rende, Cosenza, Italy}
\vskip 0.3cm
\centerline{M. Gravina$^{***}$}
\vskip 0.3cm
\centerline{\sl Laboratoire de Physique Th\'eorique, Universit\'e de Paris-Sud 
11, B\^atiment 210}
\centerline{\sl 91405 Orsay Cedex France}
\centerline{\sl and Department of Physics, University of Cyprus,
P.O. Box 20357, Nicosia, Cyprus}
\vskip 0.3cm
\centerline{A. Papa$^{**}$}
\vskip 0.3cm
\centerline{\sl Dipartimento di Fisica, Universit\`a della Calabria,}
\centerline{\sl and Istituto Nazionale di Fisica Nucleare, Gruppo collegato
di Cosenza}
\centerline{\sl I-87036 Arcavacata di Rende, Cosenza, Italy}
\vskip 0.6cm

\begin{abstract}
We investigate the critical properties of the two-dimensional $Z(5)$ vector 
model. For this purpose, we propose a new cluster algorithm, valid for
$Z(N)$ models with odd values of $N$.
The two-dimensional $Z(5)$ vector model is conjectured to exhibit two phase 
transitions with a massless intermediate phase. We locate the position of 
the critical points and study the critical behavior across both phase 
transitions in details. In particular, we determine various critical indices 
and compare the results with analytical predictions.
\end{abstract}
 
\vfill
\hrule
\vspace{0.3cm}
{\it e-mail addresses}: 

$^*$oleg@bitp.kiev.ua, $^{**}$cortese,fiore,papa@cs.infn.it,
$^{***}$gravina@ucy.ac.cy

\newpage

\section{Introduction}
\label{intro}

The Berezinskii-Kosterlitz-Thouless (BKT) phase transition is known to take 
place in a variety of two-dimensional ($2D$) systems: certain spin models,
two-dimensional Coulomb gas, sine-Gordon model, Solid-on-Solid model, etc.,
the most popular and elaborated case being the two-dimensional $XY$ 
model~\cite{berezin,kosterlitz1,kosterlitz2}.
There are several indications that this type of phase transition is not a rare 
phenomenon in gauge models at finite temperature: one can argue that in some 
three-dimensional lattice gauge models the deconfinement phase transition is 
of BKT type as well. 
Here we are going to study an example of lattice spin model where this type 
of transition exhibits itself, namely the $2D$ $Z(N)$ spin model, also 
known as vector Potts model. 

Consider a $2D$ lattice $\Lambda = L^2$ with linear extension $L$ and impose 
periodic boundary conditions on spin fields in both directions. 
The partition function of the model can be written as
\begin{equation}
Z(\Lambda, \beta ) =\left[ \prod_{x\in \Lambda} 
\frac{1}{N} \sum_{s(x)=0}^{N-1} \right ] \left[  \prod_{x\in\Lambda} 
\prod_{n=1,2} Q \left ( s(x)-s(x+e_n) \right) \right] \;.
\label{PFZNdef}
\end{equation}
In the standard formulation the most general $Z(N)$-invariant Boltzmann weight 
with $N-1$ different couplings is 
\begin{equation}
Q(s)\ =\ \exp\left[\sum_{k=1}^{N-1} \beta_k \cos\frac{2\pi k}{N}s \right]\ .
\label{Qpgen}
\end{equation}
In the Villain formulation the Boltzmann weight reads instead
\begin{equation}
Q(s) \ = \ \sum_{m=-\infty}^{\infty}
\exp\left[-\frac{1}{2}\beta\left(\frac{2\pi}{N}s+2\pi m\right)^2 \right] \ .
\label{QpVillain}
\end{equation}

Some details of the critical behavior of $2D$ $Z(N)$ spin models are well known
-- see the review in Ref.~\cite{Wu}. The $Z(N)$ spin model in the Villain 
formulation~(\ref{QpVillain}) has been studied analytically in 
Refs.~\cite{elitzur,savit,kogut,nienhuis,kadanoff}. It was shown that the 
model has at least two phase transitions when $N\geq 5$. The intermediate 
phase is a massless phase with power-like decay of the correlation function. 
The critical index $\eta$ has been estimated both from the renormalization 
group (RG) approach of the Kosterlitz-Thouless type and from the weak-coupling 
series for the susceptibility. 
It turns out that $\eta(\beta^{(1)}_{\rm c})=1/4$ at the transition point from 
the strong coupling (high-temperature) phase to the massless phase, {\it i.e.} 
the behavior is similar to that of the $XY$ model. At the transition point 
$\beta^{(2)}_{\rm c}$ from the massless phase to the ordered low-temperature 
phase one has $\eta(\beta^{(2)}_{\rm c})=4/N^2$. 
A rigorous proof that the BKT phase transition does take place, and so that 
the massless phase exists, has been constructed in Ref.~\cite{rigbkt} for both 
Villain and standard formulations (with one non-vanishing coupling $\beta_1$).
Monte Carlo simulations of the standard version with $N=6,8,12$ were performed 
in Ref.~\cite{cluster2d}. Results for the critical index $\eta$ agree well 
with the analytical predictions obtained from the Villain formulation of the 
model.

In this paper we thoroughly investigate the case $N=5$, the lowest number 
where the BKT transition is expected. Precisely, we concentrate on the 
standard formulation (\ref{Qpgen}) with one non-zero coupling $\beta_1$. 
The motivation of our study is three-fold: 
\begin{enumerate}
\item 
to compute critical indices at the transition points, which could serve 
as checking point of universality; 
\item
to shed light on the discrepancy in the literature concerning the $Z(5)$ model;
\item 
to develop and test a new version of Monte Carlo cluster algorithm valid for 
odd values of $N$.
\end{enumerate}

The first motivation is related to the study of the finite-temperature transitions 
in $3D$ $Z(N)$ and $SU(N)$ lattice gauge theory (LGT).
It is expected that in $3D$ $Z(N)$ LGT a deconfinement 
phase transition takes place at finite temperature. There is no precise 
statement about the order of the phase transition, but presumably it is of the 
BKT type if $N>4$. If it is the case, the Svetitsky-Yaffe 
conjecture~\cite{svetitsky} implies that the $3D$ $Z(N)$ LGT is in the 
universality class of $2D$ vector Potts model. 
Moreover, it can be proven that, in the strong coupling region with respect to 
the spatial coupling, the $3D$ $Z(N)$ LGT reduces to a $2D$ $Z(N)$ model with 
the general Boltzmann weight~(\ref{Qpgen}), and such that 
$\beta_1 = \beta_4 \gg \beta_2 = \beta_3$, for $N=5$. Here, $\beta_k$ are 
effective couplings 
which depend on the gauge coupling and the temporal extension $N_t$. Thus, 
our $Z(5)$ model represents a good approximation to $3D$ $Z(5)$ LGT in this 
region.

Next, let $W(x)\in SU(N)$ and consider the following effective action in $2D$
\begin{equation}
S_{\rm eff} \ = \ \sum_{x,n}  \ {\rm Tr}W(x){\rm Tr}W^{\dagger}(x+e_n) 
+ {\rm c.c.}\; .
\label{sunaction}
\end{equation}
The effective action~(\ref{sunaction}) can be regarded as the simplest 
effective model for the Polyakov loop which can be derived in the strong 
coupling region of $3D$ $SU(N)$ LGT at finite temperature. It possesses 
$Z(N)$ global symmetry and thus may well exhibit the BKT transitions which belong 
to the universality class of the corresponding vector Potts model. 
Therefore, our investigation here can be viewed as a preliminary step in 
studying deconfinement phase transition in $3D$ $Z(N)$ and $SU(N)$ LGTs. 

The second motivation reflects the fact that many features of the critical behavior 
of the $Z(5)$ model are not reliably established. Moreover, there are certain 
discrepancies even in determining the nature of the phase transition, 
{\it i.e.} whether the phase transition is of BKT type or not. 

Let us briefly summarize the present state of affairs.
\begin{itemize} 
\item 
The rigorous proof of the massless phase existence in $2D$ $Z(N)$ models 
utilizes methods which do not allow to establish the exact value of $N$ 
above which the BKT phase transition exists~\cite{rigbkt}. 

\item

Reliable analytical calculations can only be performed with the Villain 
formulation (\ref{QpVillain}); the RG study of Ref.~\cite{elitzur} predicts 
that the massless phase exists for all $N>4$.  

\item 
Some information on the phase structure of the general $Z(N)$ spin models 
can be obtained through the duality transformations (see, e.g.~\cite{ukawa}). 
These transformations cannot be used to establish the position of the critical 
points in the $Z(5)$ model~\cite{Wuz5}. However, duality transformations relate
the two critical points and thus can be used to verify the accuracy of 
numerical data. Moreover, one can predict an approximate phase diagram and 
argue that the massless phase and the BKT transition exist for $N=5$ in a 
certain region of the parameter space~\cite{Cardy,Domany}. Important in this 
context is the rigorous proof of the existence of the massless phase, 
constructed in Ref.~\cite{Cardy} for the generalized Villain formulation. 
This generalized formulation contains the vector Potts model defined
in~(\ref{Qpgen}) with one non-zero coupling $\beta_1$ as a particular case. 

For completeness we mention that in Refs.~\cite{Wyld} it was suggested 
that there is only one first order phase transition in $Z(5)$ model. This, 
however, contradicts the rigorous results of~\cite{Cardy}.  

\item 
An analytical prediction for the critical index $\eta$ has been obtained for 
the Villain formulation in Ref.~\cite{elitzur}: 
$\eta(\beta^{(1)}_{\rm c})=1/4$ and  
$\eta(\beta^{(2)}_{\rm c})=0.16$ (for $N=5$). The situation remains unclear 
for the index $\nu$ which governs the behavior of the correlation length. 
Normally, the value of $\nu$ can be estimated from the solution of the RG
equations like in the $XY$ model~\cite{kosterlitz2}. 
The analytical solution of the system of RG equations for $Z(N)$ vector models 
is unknown (see Ref.~\cite{elitzur}). Therefore, strictly speaking, 
there are no strong theoretical arguments indicating that $\nu=1/2$,
similarly to the $XY$ model. Moreover, the strong coupling expansion of the
$Z(5)$ model combined with Pad\'e approximants predicts that 
$\nu\approx 0.22$~\cite{elitzur}.  

\item 
Monte Carlo simulations of the $Z(N)$ model have been performed in 
Refs.~\cite{cluster2d,Pfeifer,BM10,BMK09}. Results of Ref.~\cite{Pfeifer}, 
though obtained on rather small lattices, indicate that the BKT transition 
takes place in models with $N\geq 8$. This contradicts the results 
of~\cite{cluster2d,BMK09} which well agree with the BKT behavior for $N=6$. 

However, most recent simulations of the helicity modulus in the $Z(5)$ 
model at $\beta^{(1)}_{\rm c}$ do not agree with what is expected 
at the BKT transition~\cite{BM10}. 
Namely, at the critical point the helicity modulus 
is expected to jump discontinuously to zero, and the jump is observed for 
$Z(6)$ model, while in $Z(5)$ the helicity modulus stays small but 
non-vanishing in the high-temperature region $\beta<\beta^{(1)}_{\rm c}$. 

\end{itemize}

Still, it remains unclear how this behavior of the helicity modulus 
influences other features of the BKT transition. 
The key feature of the massless BKT phase in $Z(N)$ models is the enhancement 
of the discrete symmetry of the Hamiltonian: the symmetry of the ground state 
in the intermediate phase is rather $U(1)$ than $Z(N)$~\cite{rigbkt}. 
This can be seen in a characteristic distribution of the complex magnetization,
in the power-like decay of the correlation functions in the massless phase, 
in the vanishing of the beta-function, etc..
Also, on the basis of the universality one could conjecture that the critical 
indices in the $Z(5)$ model are the same both in the standard and Villain 
formulations. We are not aware of any numerical calculations of these 
quantities in $Z(5)$. Here we would like to fill this gap by computing 
various quantities and extracting critical indices at both transitions. 
Among the main quantities calculated in this paper there are Binder cumulants.
As RG invariant quantities, Binder cumulants are very 
useful in locating the critical couplings and determining the nature of 
the phase transition. Indeed, the computation of Binder cumulants proved to be 
very efficient in studying BKT transitions in a variety of models, like 
the $XY$ model, the discrete Gaussian model and the SOS 
model~\cite{Hasenbusch1}. 
We therefore believe that these cumulants are of great value also in the
investigation of phase transitions in $2D$ $Z(N)$ models.  
Preliminary results of our study have been presented in Ref.~\cite{lat_10}.

The paper is organized as follows: in Section~\ref{setup} we describe the 
set-up of the Monte Carlo simulation and the newly developed cluster 
algorithm; in Section~3 we introduce the observables adopted in this work and 
study the transition from the high-temperature to the massless phase; in 
Section~4 we move on to consider the transition from the massless to the 
low-temperature ordered phase; finally in Section~5 we draw our conclusions. 
In the Appendix we check the consistency of our determination of the 
critical couplings with the duality transformations. 

\section{Algorithm and numerical set-up}
\label{setup}

In this work we concentrate our attention to the model defined by 
Eqs.~(\ref{PFZNdef}) and~(\ref{Qpgen}), with only one non-zero coupling, 
$\beta_1\equiv \beta$~\footnote{All forthcoming tables and plots refer to
the case $N=5$, but we nevertheless present all definitions and formulae 
for a generic $N$.}.
This model is known in the literature also as $N$-state ferromagnetic clock 
model and is a discrete version of the continuous $XY$ (plane rotator) model. 
It consists of 2D planar spins restricted to $N$ evenly spaced directions, 
with spin interaction energy proportional to their scalar product. 

The Hamiltonian of the model is
\begin{equation}
H=-\beta\sum_{\langle ij \rangle}\cos\left(\frac{2\pi}{N}(s_{i}-s_{j})\right)
\;,\;\;\;\;\; s_i=0,1,\ldots, N-1\;,
\label{ene}
\end{equation}
with summation taken over nearest-neighbor sites. For $N=2$ this is the Ising 
model, whereas in the $N\rightarrow\infty$ limit we get the $XY$ model.

Here we develop a new algorithm, valid for odd $N$, by which an accurate 
numerical study of the model can be performed for $N=5$, {\it i.e.} 
the smallest $N$ value for which the phase structure described 
in the Introduction holds.

Here are the steps of our cluster algorithm for the update of a spin 
configuration $\{s_i\}$:
\begin{itemize}
\item choose randomly $n$ in the set $\{0,1,2,\ldots,N-1\}$

\item build a cluster configuration according to the following probability
of bond activation between neighboring sites $ij$
\[
p_{ij}= \left\{
\begin{array}{cl}
1-\exp(-2 \beta\ \alpha_{i}\alpha_{j})\;\;& {\rm if}\;\alpha_{i}\alpha_{j}>0 \\
0                                            & {\rm otherwise} \\
\end{array}\right. \;, \;\;\;\;
{\rm with} \;\; 
\alpha_k \equiv \sin\left(\frac{2\pi}{N}(s_k-n)\right)
\]

\item ``flip'' each cluster, with probability 1/2, by replacing all its spins 
according to the transformation 
\[
s_i \rightarrow {\rm mod}(-s_i+2n+N,N)\;,
\] 
which amounts to replacing each spin $s_i$ in a cluster by the spin $s_j$ for 
which $\alpha_j=-\alpha_i$; equivalently, if the spins are mapped into the $N$ 
roots of unity in the complex plane, the above replacement means flipping the 
component of each spin transverse to the direction identified by $n$.

\end{itemize}
It is easy to prove that this cluster algorithm fulfills the detailed balance.

We have tested the efficiency of the cluster algorithm against the standard 
heat-bath algorithm. On a lattice with $L=64$ we simulated the model with $N=5$
and determined the autocorrelation time $\tau$ of three observables: the 
energy, defined in Eq.~(\ref{ene}), the magnetization $M_L$ and the population 
$S_L$, to be defined below. We considered three $\beta$ values (0.80, 1.10 and 
1.50) lying, respectively, in the high-temperature, massless and 
low-temperature phase of the model. Results are summarized in Table~\ref{tau}, 
whose last two columns give also the number of sweeps needed to reach thermal 
equilibrium and the computer time to collect a 50k statistics. 

\begin{table}[ht]
\centering\caption[]{Cluster versus heat-bath in $Z(5)$ on a $64^2$ lattice
at three values of $\beta$: autocorrelation time $\tau$ for three observables 
(energy, magnetization $M_L$ and population $S_L$), number of thermalization 
sweeps and computer time for 50k updates.}
\vspace{0.4cm}
\begin{tabular}{|c|lccc|cc|}
\hline
&           & Energy    & $M_L$      & $S_L$     & Thermalization & Time    \\
\hline
& cluster   & 5.135(82) &  1.528(38) & 1.494(39) & $\sim 10$      & 46.69 s \\
\raisebox{1.5ex}{$\beta=0.80$} 
& heat-bath & 5.43(33)  & 12.83(27)  & 12.65(28) & $\sim 100$     &  1290 s \\
\hline
& cluster   & 7.36(11)  & 5.56(10)   &  7.18(11) & $\sim 10$      & 45.14 s \\
\raisebox{1.5ex}{$\beta=1.10$} 
& heat-bath & 10.11(48) & 48.3(4.8)  & 60.6(6.1) & $\sim 1000$    &  1194 s \\
\hline
& cluster   & 8.97(17)  & 8.71(17)   &  8.84(17) & $\sim 100$     & 42.60 s \\
\raisebox{1.5ex}{$\beta=1.50$} 
& heat-bath & 2.38(12)  & 3.73(15)   &  3.78(13) & $\sim 6500$    &  1064 s \\
\hline
\end{tabular}
\label{tau}
\end{table}

At $\beta$=0.80 and 1.10 the autocorrelation time in the cluster algorithm is 
lower than in the heat-bath for the energy and much lower for magnetization and
population. At $\beta=1.50$, deep in the low-temperature ordered phase, $\tau$ 
is systematically higher in the cluster than in the heat-bath. This is a 
consequence of the lowering of the bond activation probability for increasing 
$\beta$. This drawback, however, is compensated by the higher simulation speed,
with respect to the heat-bath algorithm. Moreover since the two transitions in 
the $2D$ $Z(5)$ model are rather close (see below), there is no doubt that the 
cluster algorithm is strongly preferable.

\begin{figure}[tb]
\vspace{0.5cm}
\centering
\includegraphics[scale=0.27]{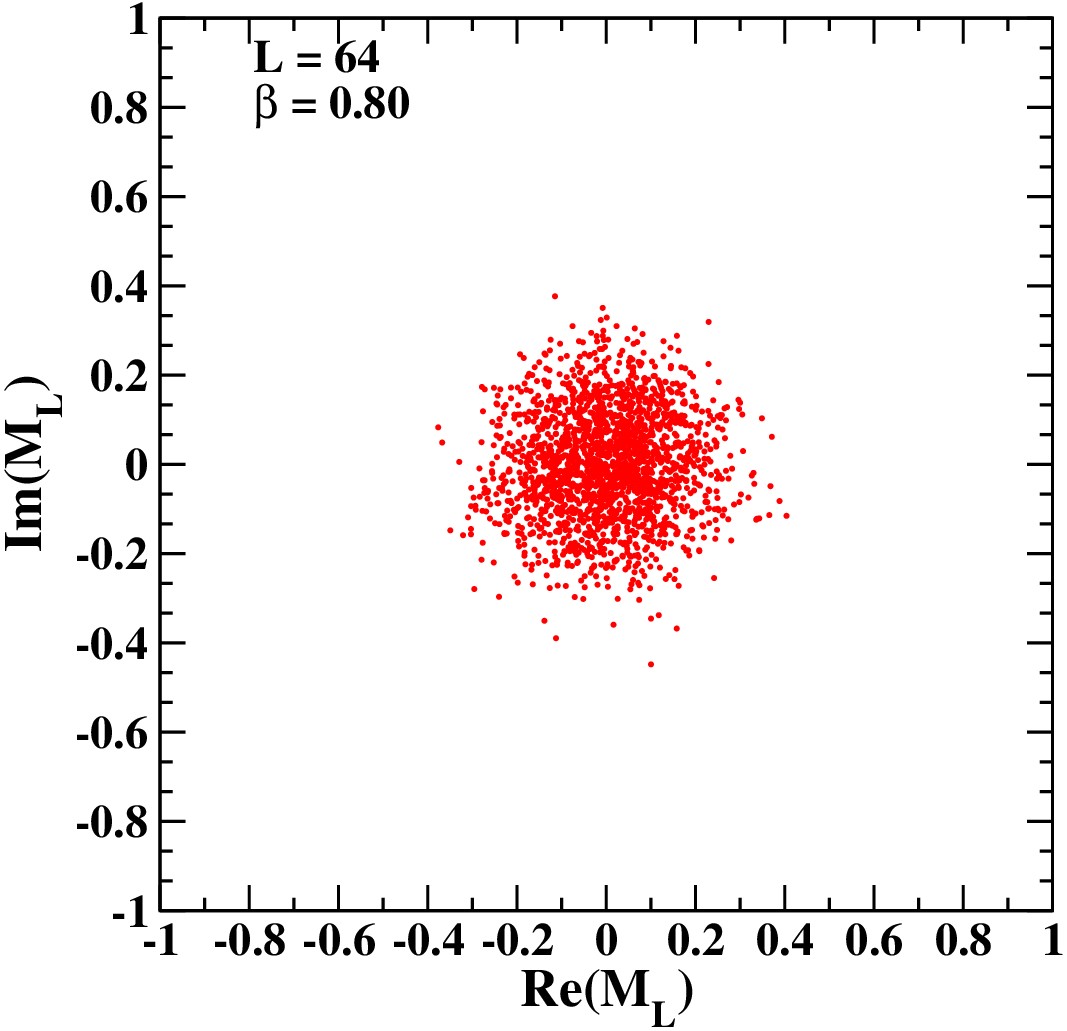} \hspace{0.2cm}
\includegraphics[scale=0.27]{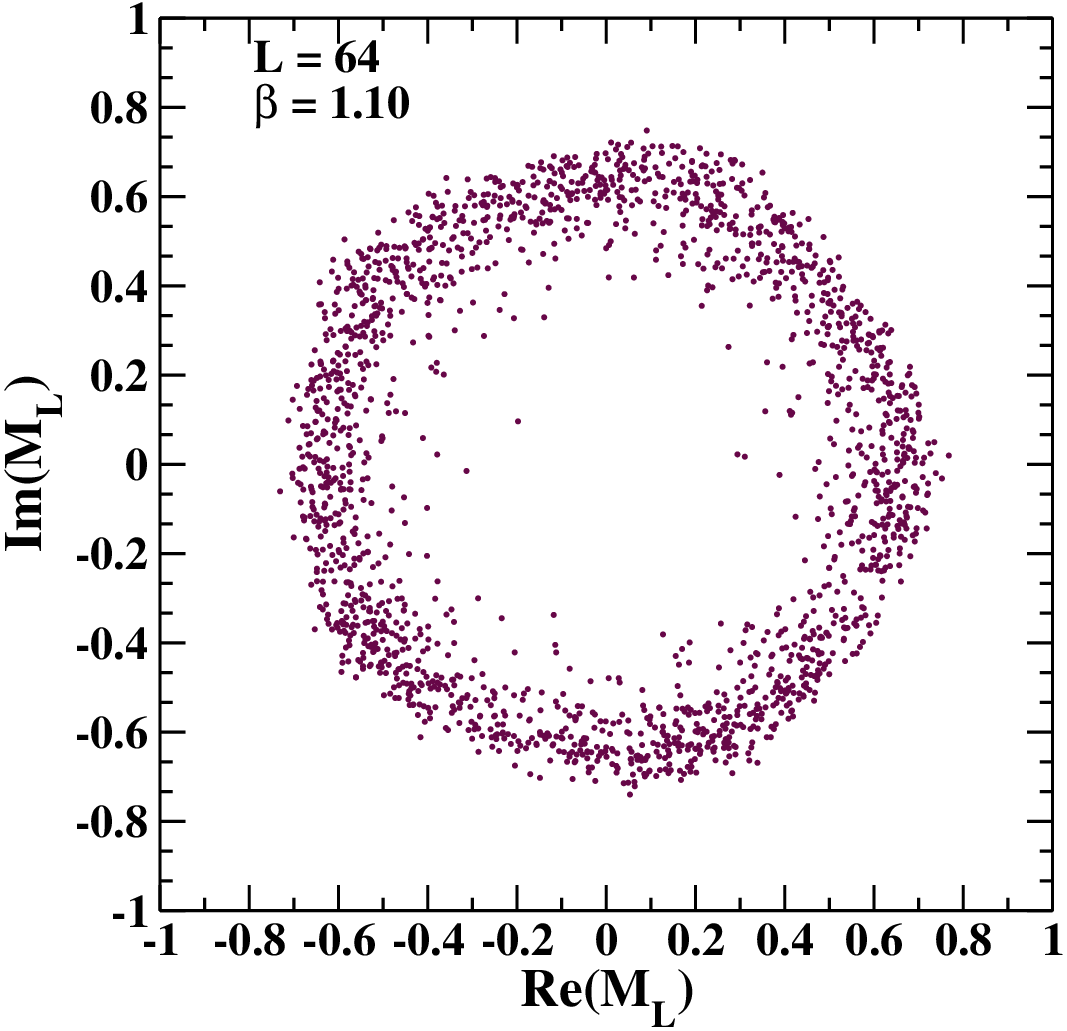} \hspace{0.2cm}
\includegraphics[scale=0.27]{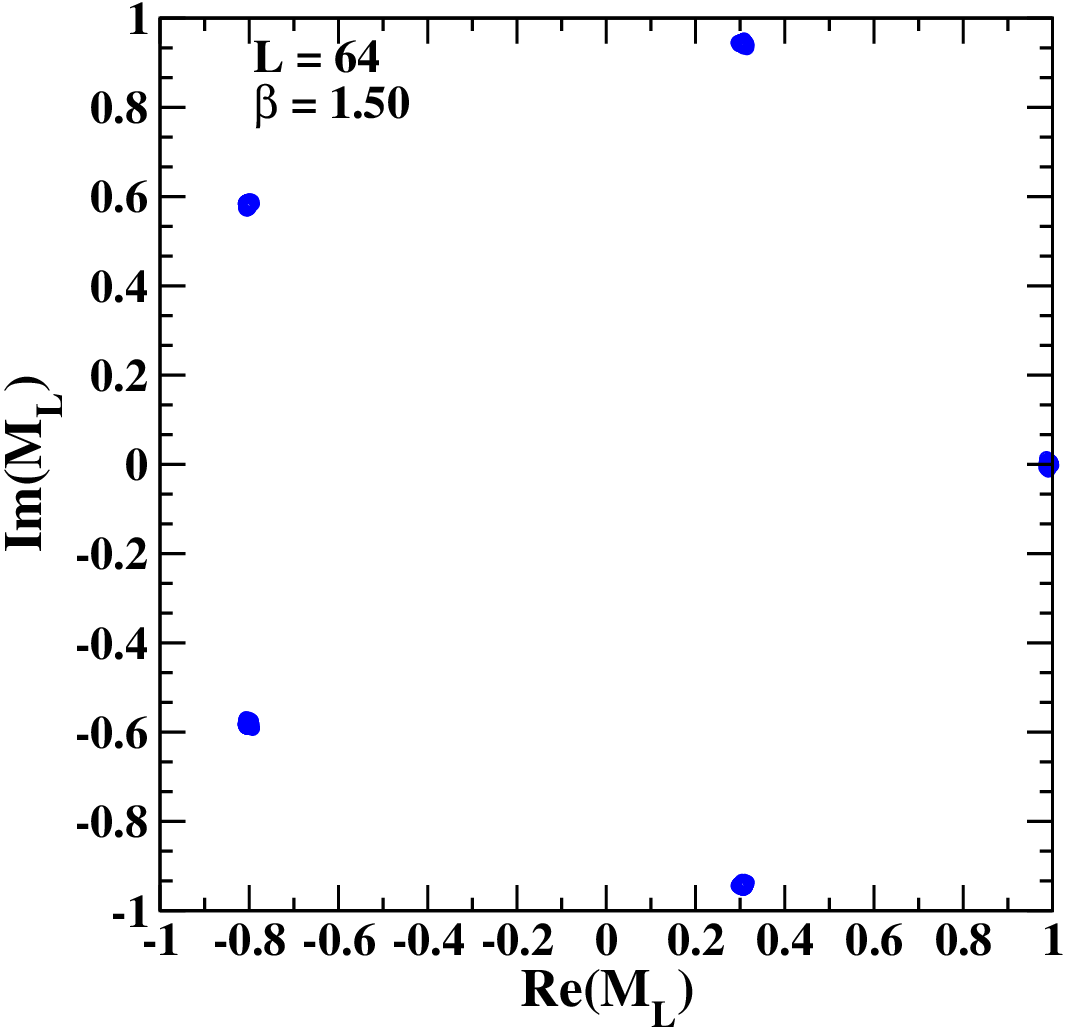}
\caption{(Color online) Scatter plot of the complex magnetization $M_L$ at 
$\beta=0.80, 1.10, 
1.50$ in $Z(5)$ on a $64^2$ lattice.}
\label{fig:scatter}
\end{figure}

The improvement brought along by the cluster algorithm becomes more visible 
when the dynamical critical exponent $z$ is considered, defined as $\tau \sim 
\xi^z$, where $\xi$ is the correlation length. We have evaluated $z$ in the 
$2D$ $Z(5)$ model on lattices with $L=16,32,64,128,256,384,512$ at both 
transition points, using the autocorrelation time of the magnetization $M_L$. 
Since at both points the correlation length diverges, the expected scaling law 
becomes $\tau \sim L^{z}$. We got in all cases that $\tau$ keeps almost 
constant at $\approx 7$, thus implying $z\simeq 0$, {\it i.e.} no critical 
slowing down.

The three phases exhibited by the $2D$ $Z(5)$ spin model can be characterized
by means of two observables: the {\em complex magnetization} $M_L$ and the
{\em population} $S_L$.

The complex magnetization is given by 
\begin{equation}
M_L=\frac{1}{L^2}\sum_i \exp\left(i\frac{2\pi}{N} s_i\right)
\equiv |M_L|e^{i\psi}\;.
\label{magn}
\end{equation}
In Fig.~\ref{fig:scatter} we show the scatter plot of $M_L$ on a lattice with 
$L=64$ in $Z(5)$ at three values of $\beta$, each representative of a different
phase: $\beta=0.80$ (high-temperature, disordered phase), $\beta=1.10$ 
(BKT massless phase) and $\beta=1.50$ (low-temperature, ordered phase).
As we can see we pass from a uniform distribution (low $\beta$) to a ring 
distribution (intermediate $\beta$) and finally to five isolated spots 
(high $\beta$).

\begin{figure}[tb]
\centering
\includegraphics[scale=0.31]{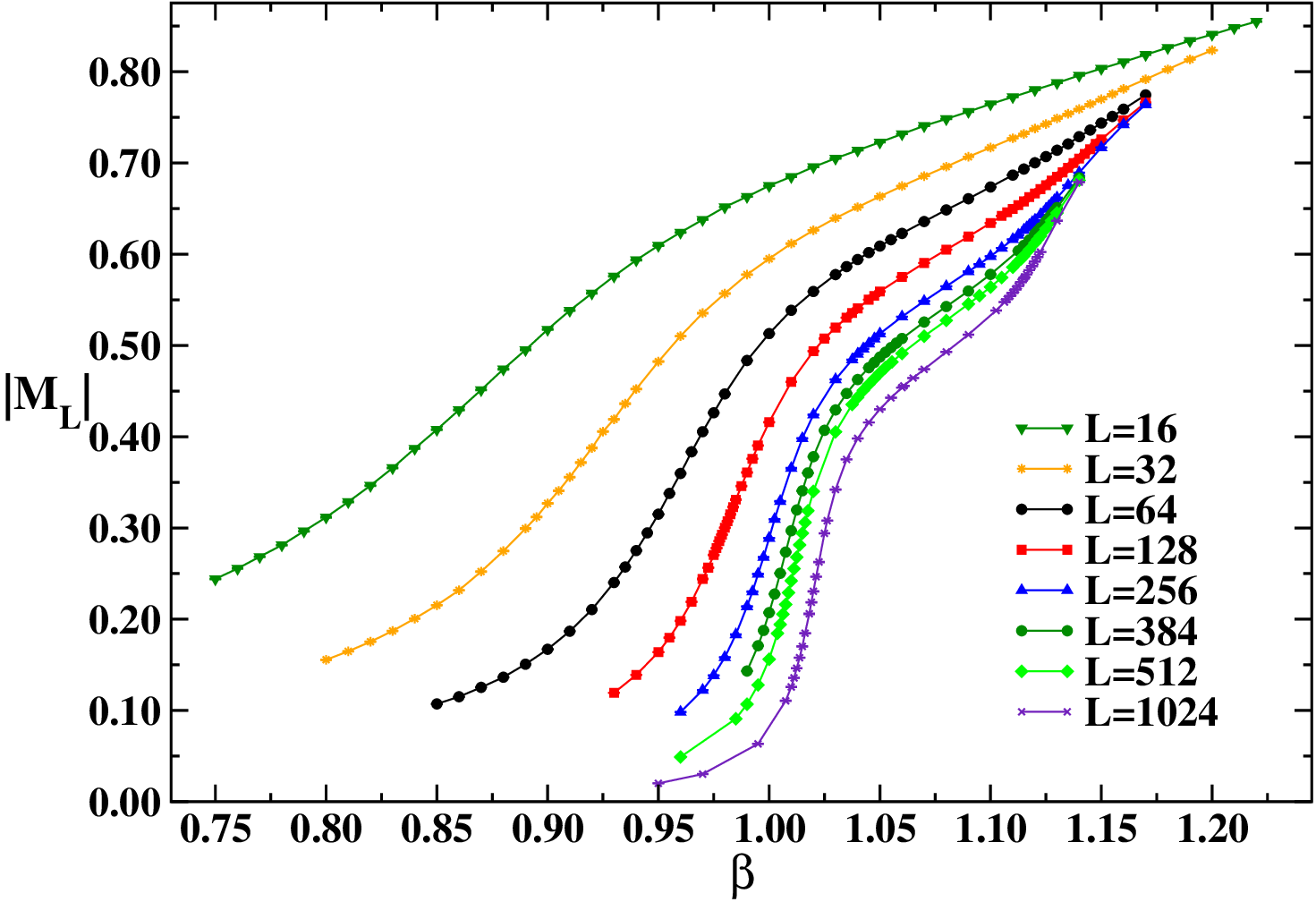} \hspace{0.2cm}
\includegraphics[scale=0.31]{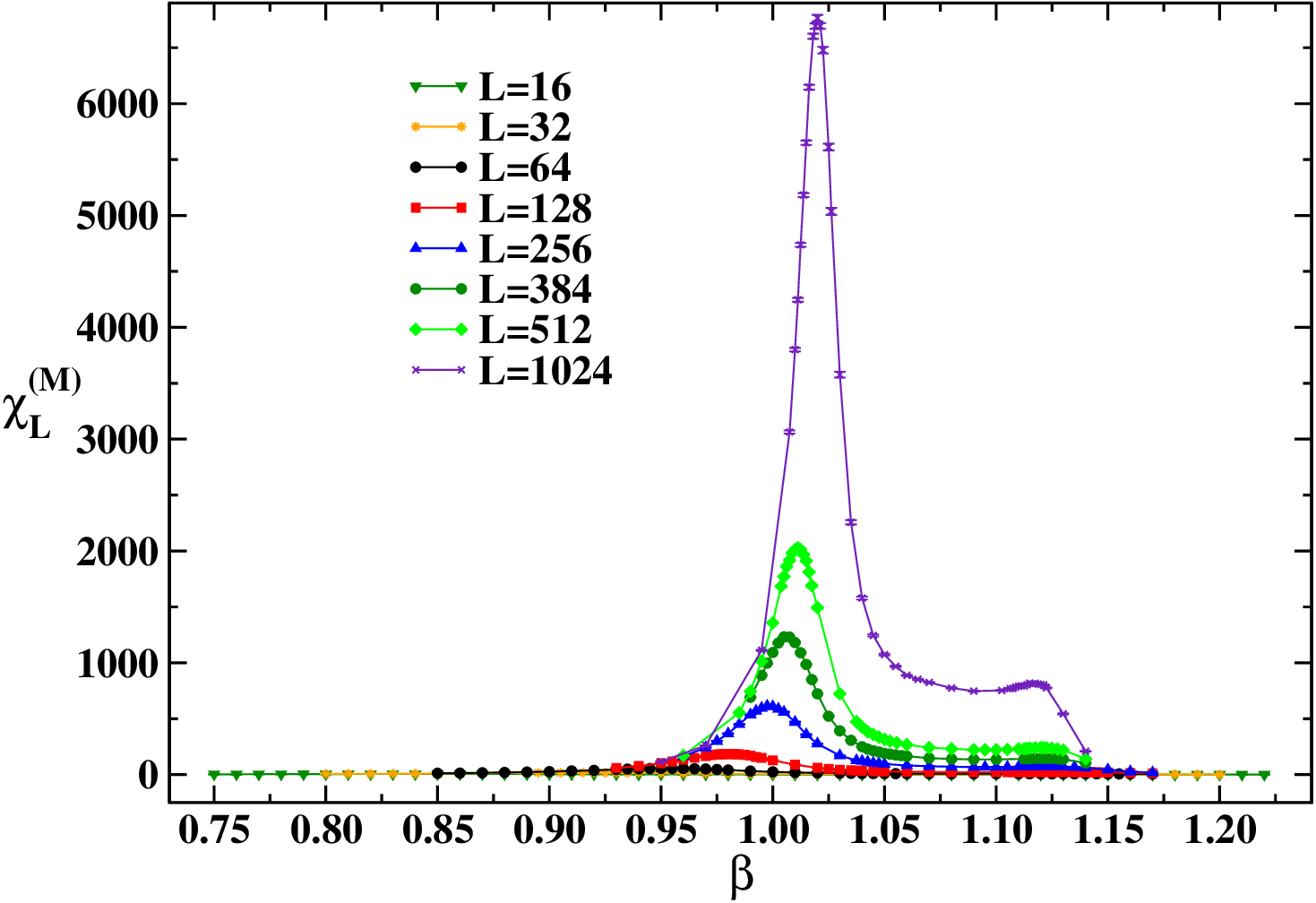}
\caption{(Color online) Behavior of $|M_L|$ (left) and of its susceptibility 
$\chi^{(M)}_L$ (right) versus $\beta$ in $Z(5)$ on lattices with $L$ ranging 
from 16 to 1024.}
\label{fig:magn}
\end{figure}

The naive average of the complex magnetization gives constantly zero, 
therefore $M_L$ is not an order parameter. An observable to detect
the transition from one phase to the other is instead the absolute 
value $|M_L|$ of the complex magnetization. In Fig.~\ref{fig:magn} we show
the behavior of $|M_L|$ and of its susceptibility,
\begin{equation}
\chi^{(M)}_L=L^2 (\langle |M_L|^2 \rangle - \langle |M_L| \rangle ^2)\;,
\label{susc_magn}
\end{equation}
in $Z(5)$ on lattices with $L$ ranging from 16 to 1024 over a wide interval of 
$\beta$ values. On each lattice the susceptibility $\chi^{(M)}_L$ clearly 
exhibits two peaks, the first of them, more pronounced than the other, 
identifies the pseudocritical coupling $\beta^{(1)}_{\rm pc}(L)$ at which 
the transition from the disordered to the massless phase occurs, whereas the 
second corresponds to the pseudocritical coupling $\beta^{(2)}_{\rm pc}(L)$ 
of the transition from the massless to the ordered phase. It is evident from 
Fig.~\ref{fig:magn} that $|M_L|$ is particularly sensitive to the first 
transition, thus making this observable the best candidate for studying its 
properties.

\begin{figure}[tb]
\vspace{0.4cm}
\centering
\includegraphics[scale=0.31]{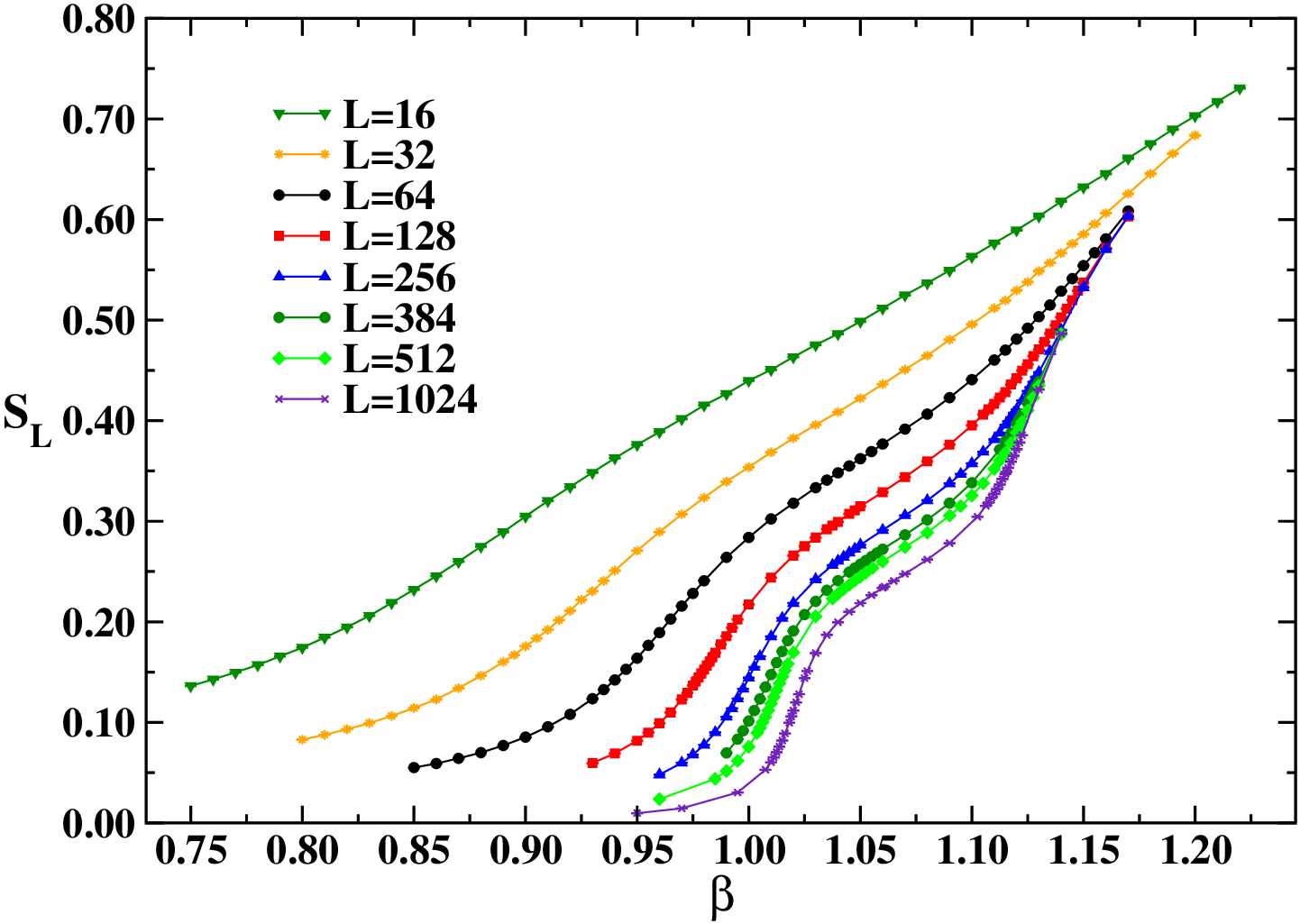} \hspace{0.2cm}
\includegraphics[scale=0.31]{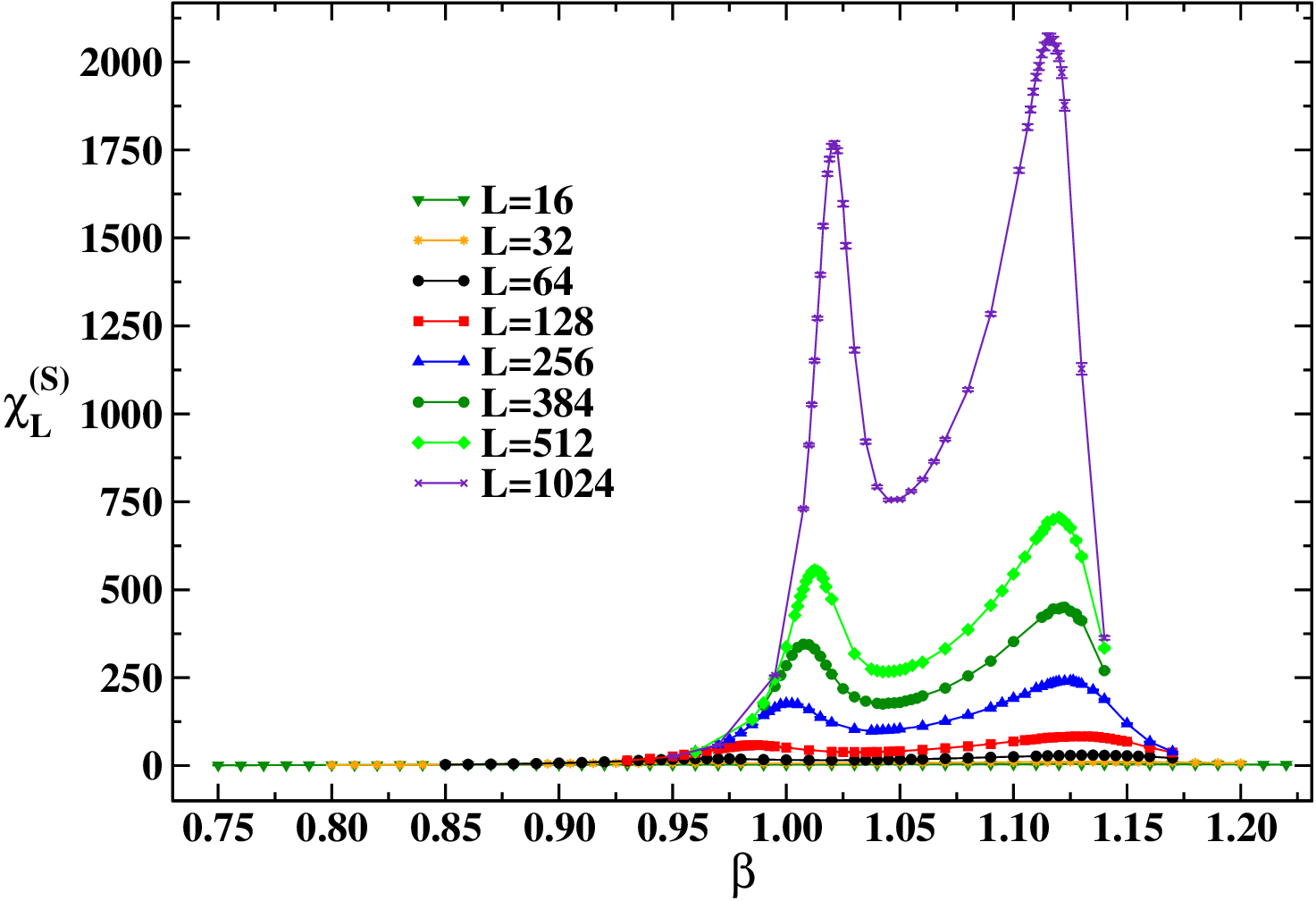}
\caption{(Color online) Behavior of $S_L$ (left) and of its susceptibility $\chi^{(S)}_L$ 
(right) versus $\beta$ in $Z(5)$ on lattices with $L$ ranging from 16 to 1024.}
\label{fig:pop}
\end{figure}

As order parameter to better detect the second transition, {\it i.e.} 
that from the massless to the ordered phase, we chose instead the 
{\it population} $S_L$, defined as 
\begin{equation}
S_L=\frac{N}{N-1}\left[\frac{\max_{i=0,N-1}(n_i)}{L^2} -\frac{1}{N}\right] \;,
\end{equation}
where $n_i$ represents the number of spins of a given configuration which are
in the state $s_i$. In a phase in which there is not a preferred spin 
direction in the system (disorder), we have $n_i\sim L^2/N$ for each index 
$i$, therefore $S_L\sim 0$. Otherwise, in a phase in which there is a 
preferred spin direction (order), we have $n_i\sim L^2$ for a given index $i$, 
therefore $S_L\sim 1$. 
In Fig.~\ref{fig:pop} we show the behavior of $S_L$ and of its susceptibility
\begin{equation}
\chi^{(S)}_L=L^2 (\langle S_L^2 \rangle - \langle S_L \rangle ^2)\;,
\label{susc_pop}
\end{equation}
in $Z(5)$ on lattices with $L$ ranging from 16 to 1024 over a wide interval 
of $\beta$ values. Again the peaks signalling the two transitions are clearly 
visible and their positions agree with Fig.~\ref{fig:magn}, but now the second 
one is more pronounced.

Other observables which have been used in this work are the following:
\begin{itemize}
\item the real part of the ``rotated'' magnetization, $M_{R}=|M_L|\cos(N\psi)$,
\item the order parameter introduced in Ref.~\cite{BMK09}, 
$m_\psi=\cos(N\psi)$,
\end{itemize}
where $\psi$ is the phase of the complex magnetization defined in 
Eq.~(\ref{magn}).

In the next two Sections we will study separately the two transitions 
of $2D$ $Z(5)$ and determine some of the related critical indices. For all 
observables considered in this work we collected typically 100k measurements, 
on configurations separated by 10 updating sweeps. For each new run the first 
10k configurations were discarded to ensure thermalization. Data analysis was 
performed by the jackknife method over bins at different blocking levels.

\section{The transition from the high-temperature to the massless phase}
\label{first_transition}

The first inflection point in the plot of the magnetization $|M_L|$ and the 
first peak in the plot of the susceptibility $\chi_L^{(M)}$ (see 
Fig.~\ref{fig:magn}) indicate the transition from the disordered to the 
massless phase. The couplings where this transition occurs (denoted as the 
pseudocritical couplings $\beta_{\rm pc}^{(1)}(L)$) have been 
determined by a Lorentzian interpolation around the peak of the susceptibility 
$\chi_L^{(M)}$. Their values are summarized in the second column of 
Table~\ref{beta1_pc}. We observe that, when the lattice size $L$ grows, 
$\beta_{\rm pc}^{(1)}(L)$ increases towards the infinite volume critical 
coupling $\beta^{(1)}_{\rm c}$ and that the susceptibility $\chi^{(M)}_{L}$ 
goes to zero less rapidly for $\beta>\beta^{(1)}_{\rm pc}$, as expected in 
the BKT scenario. 

\begin{table}[ht]
\centering
\caption[]{Values of $\beta^{(1)}_{\rm pc}$ in $Z(5)$ on $L^2$ lattices. The 
last two columns give the susceptibility $\chi^{(M)}_L$ and the 
magnetization $|M_L|$ at the infinite volume coupling constant 
$\beta^{(1)}_{\rm c}$=1.0510.}
\vspace{0.2cm}
\begin{tabular}{|c|c|c|c|}
\hline
 $L$ & $\beta^{(1)}_{\rm pc}$ & $\chi^{(M)}_L(\beta^{(1)}_{\rm c})$ 
                              & $|M_L|(\beta^{(1)}_{\rm c})$ \\
\hline
  16 & 0.8523(20)  & -           & -           \\
  32 & 0.91429(90) & -           & -           \\
  64 & 0.95373(40) & -           & -           \\
 128 & 0.98054(30) & -           & -           \\
 256 & 0.99838(20) & -           & -           \\
 384 & 1.00621(10) &  187.9(1.2) & 0.48929(13) \\
 512 & 1.01112(20) &  311.5(2.0) & 0.47181(13) \\
 640 & -           &  458.6(3.4) & 0.45918(11) \\
 768 & -           &  631.3(4.2) & 0.44863(11) \\
 896 & -           &  824.4(5.2) & 0.44004(11) \\
1024 & 1.01991(10) & 1040.0(6.9) & 0.43277(11) \\
\hline
\end{tabular}
\label{beta1_pc}
\end{table}

In order to apply the finite size scaling (FSS) program, the location of the 
infinite volume critical coupling $\beta^{(1)}_c$ is needed. In 
Refs.~\cite{3du1ft,3du1full} this was done by extrapolating the 
pseudocritical couplings to the infinite volume limit, according to a 
suitable scaling law. First order transition is ruled out by data in 
Table~\ref{beta1_pc}. Second order transition, though not incompatible with 
data in Table~\ref{beta1_pc}, is to be excluded, due to the vanishing of the 
long distance correlations combined with the clusterization property (we
will come back to this point in the last Section).
Therefore, we assume that the transition is of BKT type and adopt the scaling 
law dictated by the essential scaling of the BKT transition, {\it i.e.} 
$\xi\sim e^{bt^{-\nu}}$, which reads
\begin{equation}
\beta^{(1)}_{\rm pc}=\beta^{(1)}_{\rm c}
+\frac{A}{(\ln L + B)^{\frac{1}{\nu}}}\quad .
\label{b_pc}
\end{equation}
The index $\nu$ characterizes the universality class of the system. For 
example, $\nu=1/2$ holds for the 2$D$ $XY$ universality class. 

Unfortunately, 4-parameter fits of the data for $\beta^{(1)}_{\rm pc}(L)$ 
give very unstable results for the parameters. This led us to move to 
3-parameter fits of the data, with $\nu$ fixed at 1/2. We found, as best
fit with the MINUIT optimization code,
\[
\beta^{(1)}_{\rm c} = 1.0602(20) \;,\;\; A_1=-2.09(20) \;,\;\;B_1=0.27(18) 
\;,\;\;\chi^2/{\rm d.o.f.}=0.48 \;,\;\;L_{\rm min}=64 \;.
\]
We observe that $\beta^{(1)}_{\rm c}$ is rather far from the value of 
$\beta^{(1)}_{\rm pc}$ on the largest available lattice, thus casting some
doubts on the reliability of the extrapolation to the thermodynamic limit. 
For this reason, we turned to an independent method for the determination of 
$\beta^{(1)}_{\rm c}$, based on the use of Binder cumulants.

\begin{figure}[tb]
\centering
\includegraphics[scale=0.4]{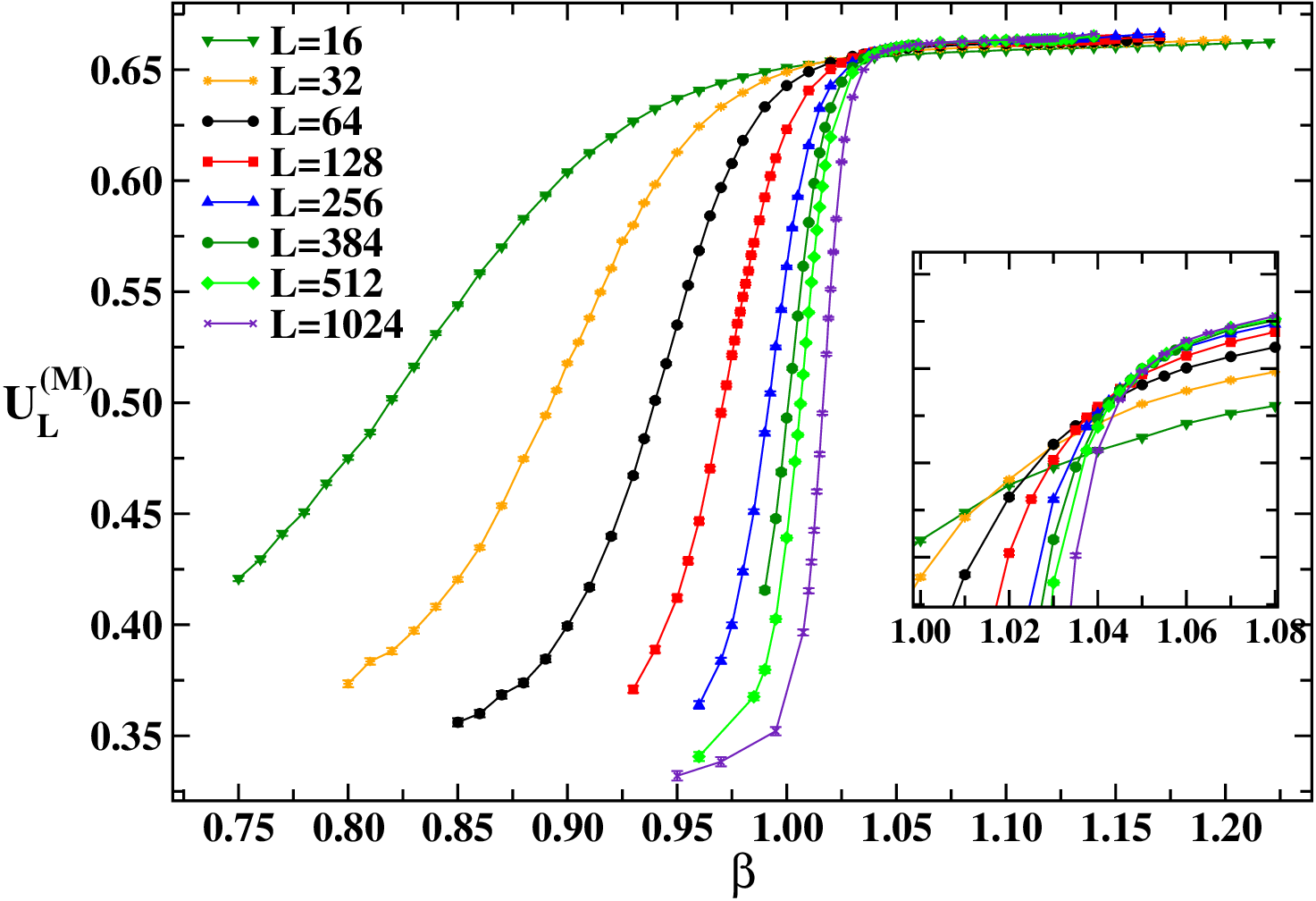}
\caption[]{(Color online) Reduced 4-th order Binder cumulant $U^{(M)}_L$ versus $\beta$ on 
lattices with $L$ ranging from 16 to 1024.}
\label{fig:binder_magn}
\end{figure}

\begin{figure}[tb]
\centering
\includegraphics[scale=0.4]{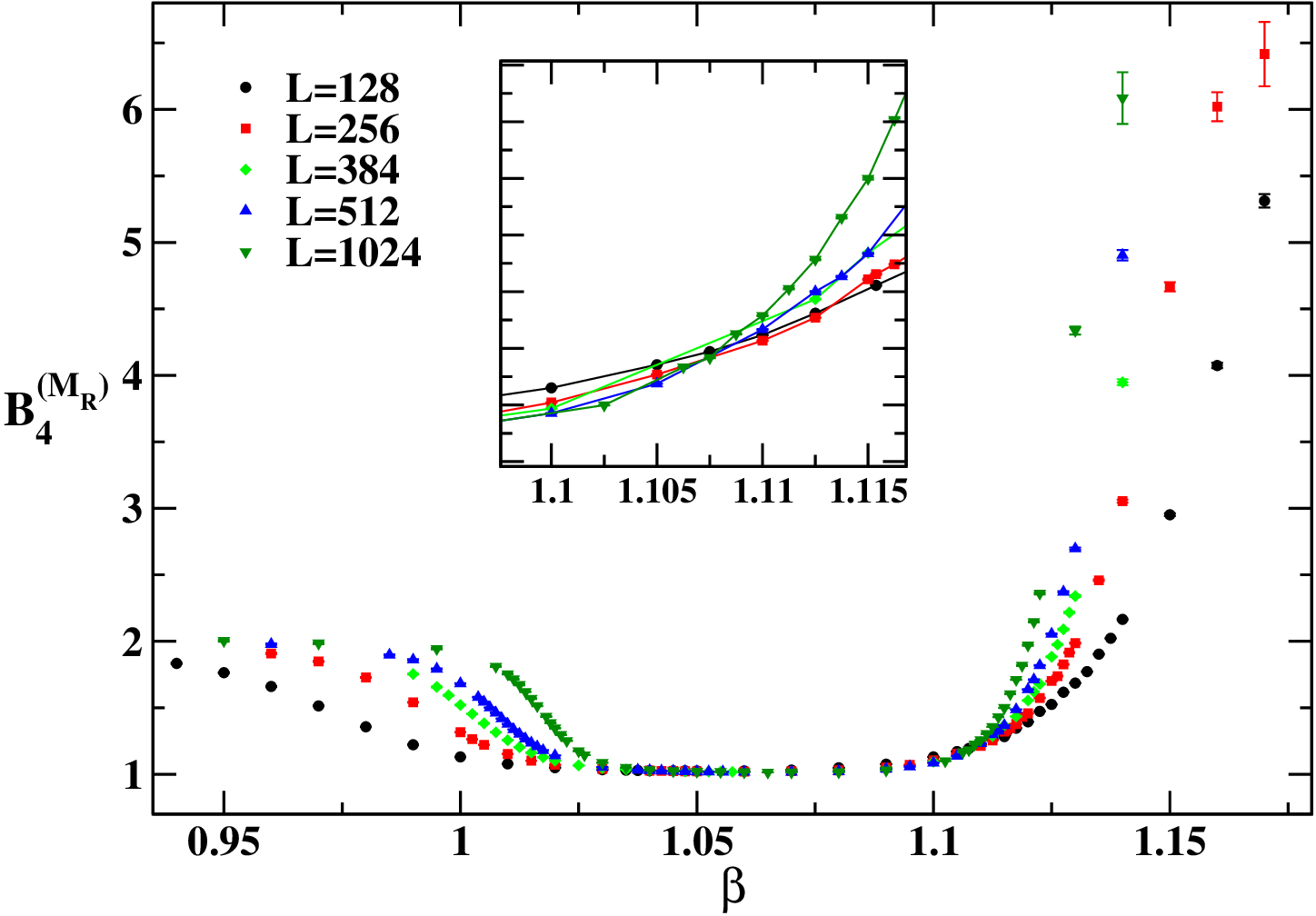}
\caption[]{(Color online) Binder cumulant $B_4^{(M_R)}$ versus $\beta$ on lattices with 
$L$ ranging from 128 to 1024.}
\label{fig:binder_MR}
\end{figure}

In particular, we considered the {\it reduced} 4-th order Binder cumulant 
$U^{(M)}_L$ defined as
\begin{equation}
U^{(M)}_L=1-\frac{\langle |M_L|^4 \rangle}{3\langle |M_L|^2 \rangle^2} \; ,
\label{binder_U}
\end{equation}
and the cumulant $B_4^{(M_R)}$ defined as
\begin{equation}
B_4^{(M_R)}=\frac{\langle |M_R-\langle M_R\rangle|^4\rangle}
{\langle |M_R-\langle M_R\rangle|^2\rangle^2}\;.
\label{binder_MR}
\end{equation}
Plots of the various Binder cumulants versus $\beta$ show that data obtained 
on different lattice volumes align on curves that cross in two 
points, corresponding to the two transitions (see Figs.~\ref{fig:binder_magn} 
and~\ref{fig:binder_MR}). We used also the {\it reduced} 4-th order Binder 
cumulant of the action which showed no crossing points nor volume-dependent
dips, thus confirming the absence of first order phase transitions.

\begin{figure}[tb]
\centering
\includegraphics[scale=0.4]{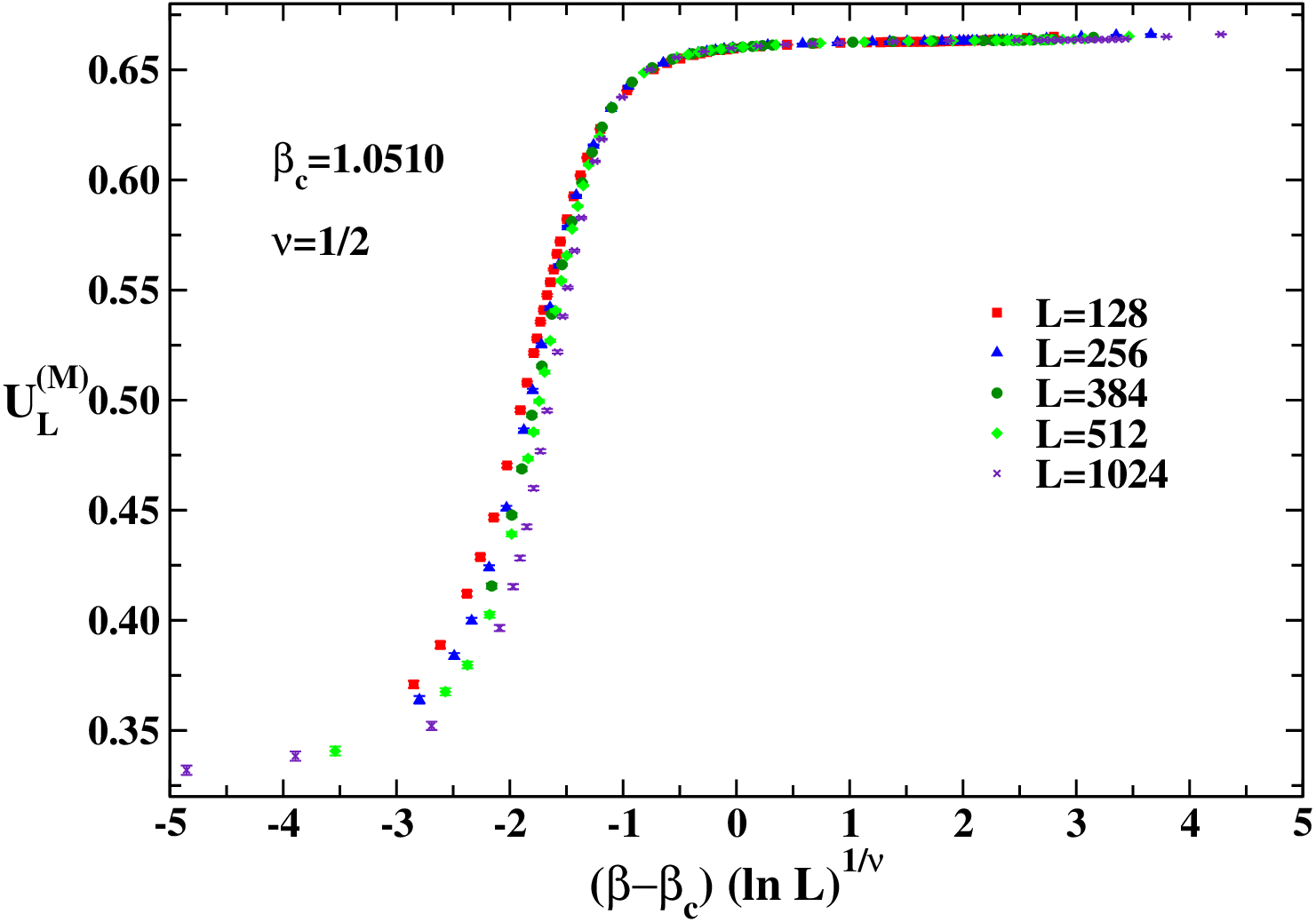}
\caption[]{(Color online) Reduced 4-th order Binder cumulant $U^{(M)}_L$ versus 
$(\beta-\beta_{\rm c})(\ln L)^{1/\nu}$, for $\beta_{\rm c}=1.0510$ and
$\nu=1/2$ on lattices with $L$ ranging from 128 to 1024.}
\label{fig:binder_rescaled}
\end{figure}

We determined the crossing point by plotting the Binder cumulants versus 
$(\beta-\beta_{\rm c})(\ln L)^{1/\nu}$, with $\nu$ fixed at 1/2, and by 
looking for the optimal overlap of data from different lattices, by the 
$\chi^2$ method (see Fig.~\ref{fig:binder_rescaled} for an example of this 
kind of plots). As a result of this analysis we arrived at the follo\-wing 
estimate: $\beta^{(1)}_{\rm c}=1.0510(10)$. We observe that 
$\beta^{(1)}_{\rm c}$ is not compatible with the infinite volume 
extrapolation of the corresponding pseudocritical couplings, thus confirming 
our previous worries about the safety of the infinite volume extrapolation 
of $\beta^{(1)}_{\rm pc}$. It should be noted, however, that a fit to 
$\beta^{(1)}_{\rm pc}(L)$ with the law~(\ref{b_pc}) and with 
$\beta^{(1)}_{\rm c}$ fixed at 1.0510 and $\nu$ fixed at 1/2 gives a good 
$\chi^2$/d.o.f., if only the three largest volumes are considered in the fit. 

We have also tested the strong coupling prediction of Ref.~\cite{elitzur} 
suggesting $\nu =0.22$. In particular, we have plotted the Binder
cumulant $U^{(M)}_L$ versus $(\beta-\beta_{\rm c})(\ln L)^{1/\nu}$, with 
$\nu$ fixed now at the candidate value 0.22. We have seen that, varying 
$\beta_{\rm c}$ on a wide interval, the overlap among curves from different 
lattices is always rather poor.

\begin{table}[ht]
\centering\caption[]{Results of the fit to the data of 
$|M_L|(\beta^{(1)}_{\rm c})$ with the scaling law~(\ref{magn_fss}) on 
$L^2$ lattices with $L\geq L_{\rm min}$.}
\vspace{0.2cm}
\begin{tabular}{|c|l|l|l|}
\hline
$L_{\rm min}$ & \hspace{0.5cm} $A$ & \hspace{0.5cm}$\beta/\nu$ & 
$\chi^2$/d.o.f. \\
\hline
384 & 1.0299(21) & 0.12508(32) & 1.3   \\
512 & 1.0294(32) & 0.12501(47) & 1.7   \\
640 & 1.0371(49) & 0.12610(71) & 0.40  \\
768 & 1.0305(89) & 0.1252(13)  & 0.021 \\
\hline
\end{tabular}
\label{M_bkt}
\end{table}

\begin{table}[ht]
\centering
\caption[]{Results of the fit to the data of 
$\chi^{(M)}_L(\beta^{(1)}_{\rm c})$ with the scaling law~(\ref{chiM_fss}) 
on $L^2$ lattices with $L\geq L_{\rm min}$.}
\vspace{0.2cm}
\begin{tabular}{|c|l|l|l|}
\hline
$L_{\rm min}$ & \hspace{0.5cm}$A$ & \hspace{0.5cm}$\gamma/\nu$ & 
$\chi^2$/d.o.f. \\
\hline
384 & 0.00586(30) & 1.7438(80) & 0.060  \\
512 & 0.00602(48) & 1.740(12)  & 0.018  \\
640 & 0.00598(81) & 1.741(20)  & 0.025  \\
768 & 0.0062(14)  & 1.735(34)  & 0.0063 \\
\hline
\end{tabular}
\label{suscetM_bkt}
\end{table}

We are now in the position to extract other critical indices and check 
therefore the hyperscaling relation. According to the standard FSS theory, in 
a $L\times L$ lattice at criticality the equilibrium magnetization $|M_{L}|$ 
should obey the relation $|M_{L}| \sim L^{-\beta / \nu}$, for sufficiently 
large $L$~\footnote{The symbol $\beta$ here denotes a critical index and not, 
obviously, the coupling of the theory. In spite of this inconvenient notation, 
we are confident that no confusion will arise, since it will be always clear 
from the context which $\beta$ is to be referred to.}. We performed a 
fit to the data of $|M_L|(\beta^{(1)}_{\rm c})$ (reported in the last column
of Table~\ref{beta1_pc}) on all lattices with size $L$ not smaller than a 
given $L_{\rm min}$ according to the scaling law
\begin{equation}
|M_{L}|=A L^{-\beta/\nu}
\label{magn_fss}
\end{equation}
and summarized our results in Table~\ref{M_bkt}.

The FSS behavior of the $\chi^{(M)}_L$ susceptibility defined in 
Eq.~(\ref{susc_magn}) is given by $\chi^{(M)}_L\sim L^{\gamma/ \nu}$, where 
$\gamma/\nu=2-\eta$ and $\eta$ is the magnetic critical index. We performed a 
fit to the data of $\chi^{(M)}_L(\beta^{(1)}_{\rm c})$ (reported in the third 
column of Table~\ref{beta1_pc}) on all lattices with size $L$ not smaller 
than a given $L_{\rm min}$ according to the scaling law
\begin{equation}
\chi^{(M)}_{L}= A L^{\gamma/\nu}
\label{chiM_fss}
\end{equation}
and summarized our results in Table~\ref{suscetM_bkt}. As we can see, for all
values of $L_{\rm min}$ considered, the value of the magnetic 
index~\footnote{The notation $^{(1)}$ in $\eta$ 
means ``at the infinite volume critical coupling of the {\it first} 
transition''.} $\eta^{(1)}\equiv 2-\gamma/\nu$ is compatible with 1/4.
Note also that the hyperscaling relation $\gamma/\nu+2\beta/\nu=d$,
where $d$ is the dimension of the system, is always satisfied within the 
statistical error.

\begin{figure}[tb]
\centering
\includegraphics[scale=0.4]{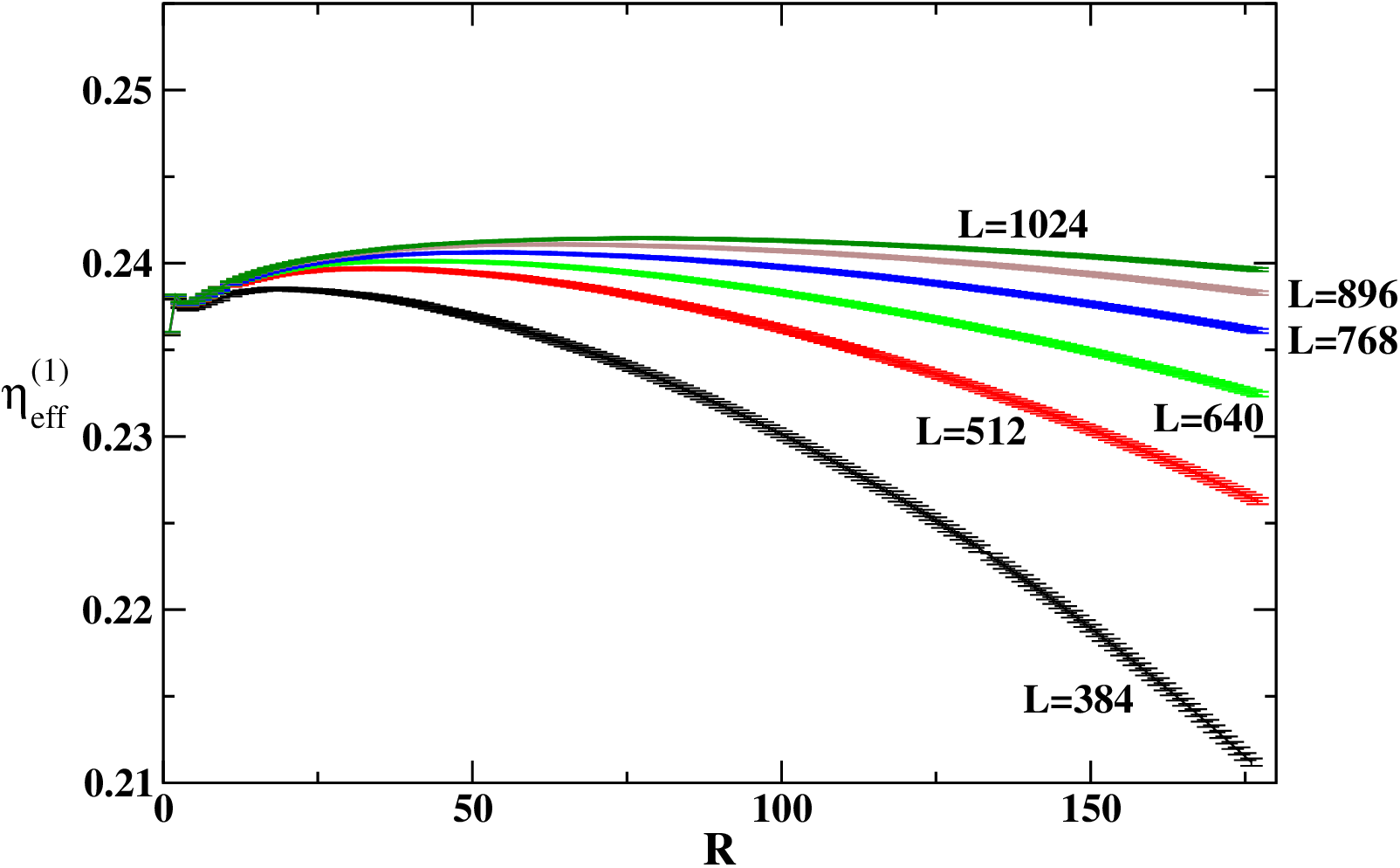}
\caption[]{(Color online) $\eta^{(1)}_{\rm eff}$ versus $R$ at 
$\beta^{(1)}_{\rm c}=1.0510$ on lattices with $L=384,512,640,768,896,1024$.}
\label{eta_eff_1}
\end{figure}

An independent determination of the magnetic index $\eta$ can be achieved by
the approach developed in Ref.~\cite{3du1ft}: an {\em effective} $\eta$ index 
is defined, through the spin-spin correlation function $\Gamma(R)$, according 
to
\begin{equation}
\eta^{(1)}_{\rm eff}(R) \equiv \frac{\ln [\Gamma (R)/\Gamma (R_0)]}
{\ln [R_0/R]} \quad ,
\label{eff_eta_def}
\end{equation}
with $R_0$ chosen equal to 10, as in Ref.~\cite{3du1ft}. This quantity is 
constructed in such a way that it exhibits a {\it plateau} in $R$ if the 
correlator obeys the law
\begin{equation}
\Gamma (R) \ \asymp \ \frac{1}{R^{\eta (T)}} \ ,
\label{corr_bkt}
\end{equation}
valid in the BKT phase, $\beta\geq\beta^{(1)}_{\rm c}$. In 
Fig.~\ref{eta_eff_1} we show the behavior of $\eta^{(1)}_{\rm eff}(R)$ at 
the infinite volume critical coupling $\beta^{(1)}_{\rm c}=1.0510$ on lattices 
with $L=384,512,640,768,896,1024$. 
It turns out that a plateau develops at small distances when $L$ increases
and that the extension of this plateau gets larger with $L$, consistently with 
the fact that finite volume effects are becoming less important. The plateau 
value of $\eta^{(1)}_{\rm eff}$ can be estimated at about 0.24.
We checked that this result is stable under variation of the parameter $R_0$.
The discrepancy with the expected value of $1/4$ can be explained
by the imperfect localization of the critical point and/or by 
the effect of logarithmic corrections~\cite{Kenna-Irving,Hasenbusch} 
that we were not able to include in our analysis.

\begin{table}[ht]
\centering
\caption[]{Values of $\beta^{(2)}_{\rm pc}$ in $Z(5)$ on $L^2$ lattices. The 
last two columns give the susceptibility $\chi^{(M_R)}_L$ and the 
rotated magnetization $M_R$ at the infinite volume coupling constant 
$\beta^{(2)}_{\rm c}$=1.1048.}
\vspace{0.2cm}
\begin{tabular}{|c|l|c|c|}
\hline
 $L$ & \hspace{0.5cm}$\beta^{(2)}_{\rm pc}$ & 
$\chi^{(M_R)}_L(\beta^{(2)}_{\rm c})$ 
                              & $M_R(\beta^{(2)}_{\rm c})$  \\
\hline
  16 & 1.1323(19)  & -           & -          \\
  32 & 1.1363(11)  & -           & -          \\
  64 & 1.13212(60) & -           & -          \\
 128 & 1.12875(66) & -           & -          \\
 256 & 1.12290(16) & -           & -          \\
 384 & 1.12103(50) &  47116(77)  & 0.1618(18) \\
 512 & 1.11912(28) &  80057(139) & 0.1575(19) \\ 
 640 & -           & 120777(229) & 0.1557(20) \\
 768 & -           & 169358(298) & 0.1517(19) \\
 896 & -           & 224879(339) & 0.1502(16) \\
1024 & 1.11596(38) & 288151(532) & 0.1473(18) \\
\hline
\end{tabular}
\label{beta2_pc}
\end{table}

\section{The transition from the massless to the low-tem\-pe\-ra\-ture 
ordered phase}
\label{second_transition}

The second inflection point in the plot of the population $S_L$ and the 
second peak in the plot of the susceptibility $\chi_L^{(S)}$ (see 
Fig.~\ref{fig:pop}) indicate the transition from the massless to the ordered 
phase. The couplings where this transition occurs (denoted as the 
pseudocritical couplings $\beta_{\rm pc}^{(2)}(L)$) have been determined by 
a Lorentzian interpolation around the peak of the susceptibility 
$\chi_L^{(S)}$. Their values are summarized in the second column of 
Table~\ref{beta2_pc}.

Available results in the literature~\cite{elitzur,cluster2d} suggest that the 
correlation length diverges according to essential scaling scenario when the 
critical point is approached from above. Our aim is to check the validity of 
this statement and then extract relevant indices characterizing the system at 
this transition. Again, first order transition is ruled out by data in 
Table~\ref{beta2_pc} (and by the aforementioned analysis of the Binder
cumulant of the action), while second order is not. We assume that a BKT 
transition is at work here and, therefore, that pseudocritical
couplings scale with $L$ according to the law~(\ref{b_pc}).
As before, 4-parameter fits of the data for $\beta^{(2)}_{\rm pc}(L)$ 
are unstable and we moved to 3-parameter fits of the data, with $\nu$ fixed at 
1/2, finding that the parameter $B_2$ turns out to be compatible with zero,
so that, in fact, a 2-parameter fit works well:
\[
\beta^{(2)}_{\rm c} = 1.1042(12)\;,\;\;A_2=0.578(41)\;,\;\;B_2=0.
\;,\;\;\chi^2/{\rm d.o.f.}=0.61\;,\;\;L_{\rm min}=128\;.
\]
Now $\beta^{(2)}_{\rm c}$ is not far from the value of $\beta^{(2)}_{\rm pc}$ 
on the largest available lattice, thus supporting the reliability of the 
extrapolation to the thermodynamic limit. 

\begin{table}[ht]
\centering
\caption[]{Results of the fit to the data of 
$\chi^{(M_R)}_L(\beta^{(2)}_{\rm c})$ with the scaling 
law~(\ref{chiM_fss}) on $L^2$ lattices with $L\geq L_{\rm min}$.}
\vspace{0.2cm}
\begin{tabular}{|c|c|c|c|}
\hline
$L_{\rm min}$ & $A$         & $\gamma/\nu$ & $\chi^2$/d.o.f. \\
\hline
384 & 0.799(11) & 1.8459(21) & 0.23 \\
512 & 0.791(17) & 1.8473(32) & 0.19 \\
640 & 0.784(28) & 1.8487(53) & 0.22 \\
768 & 0.793(50) & 1.8470(92) & 0.39 \\
\hline
\end{tabular}
\label{chiMR_second}
\end{table}

\begin{table}[ht]
\centering\
\centering\caption[]{Results of the fit to the data of 
$M_R(\beta^{(2)}_{\rm c})$ with the scaling law~(\ref{magn_fss}) on 
$L^2$ lattices with $L\geq L_{\rm min}$.}
\vspace{0.2cm}
\begin{tabular}{|c|c|c|c|}
\hline
$L_{\rm min}$ & $A$         & $\beta/\nu$  & $\chi^2$/d.o.f. \\
\hline 
384 & 0.281(26) & 0.093(14) & 0.15 \\
512 & 0.288(41) & 0.096(22) & 0.18 \\
640 & 0.322(77) & 0.112(36) & 0.12 \\
768 & 0.30(12)  & 0.102(60) & 0.19 \\
\hline
\end{tabular}
\label{MR_second}
\end{table}

\begin{figure}[tb]
\vspace{1cm}
\centering
\includegraphics[scale=0.4]{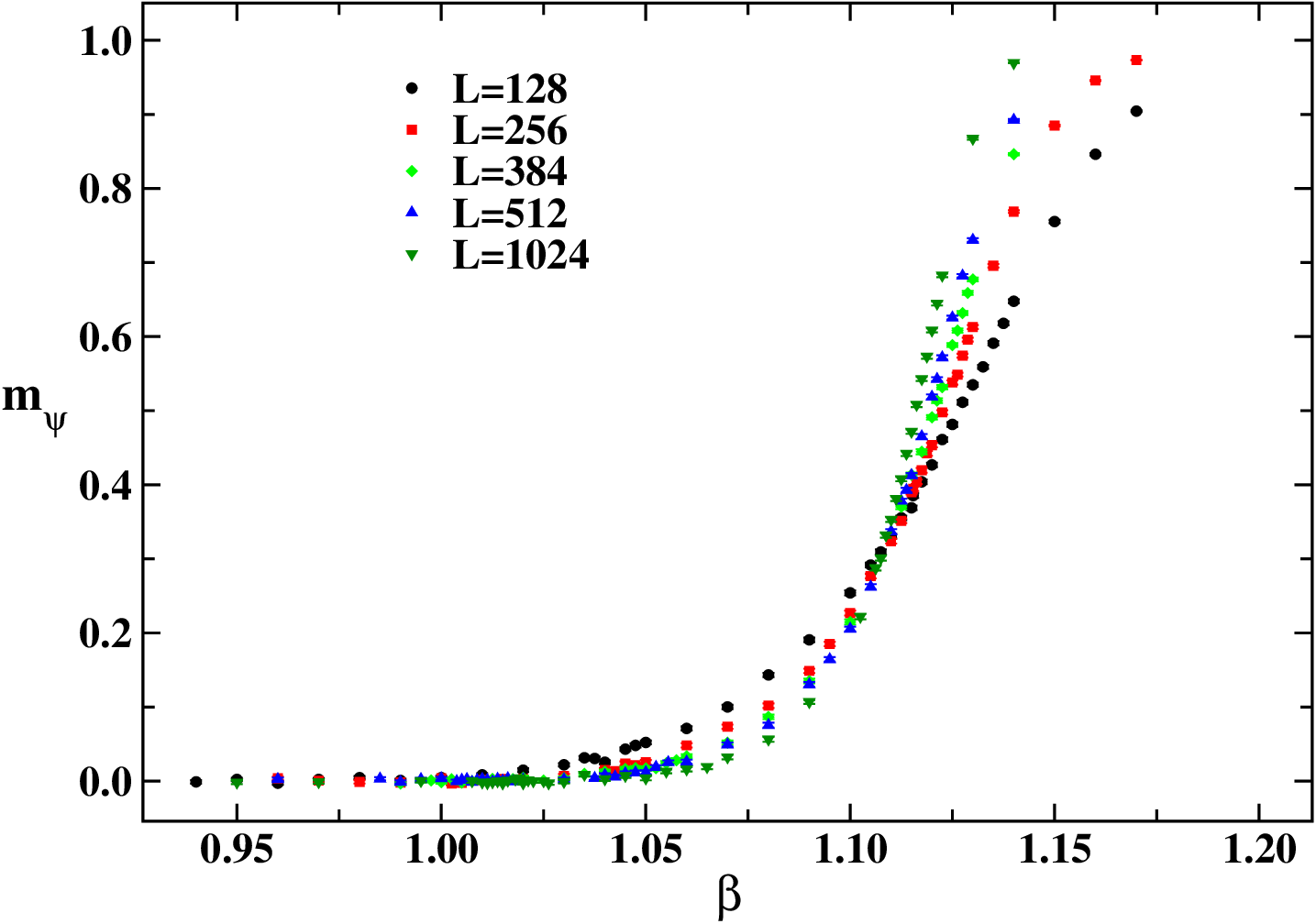} \hspace{0.2cm}
\caption{(Color online) Behavior of $m_\psi$ versus $\beta$ on lattices with 
$L$ ranging from 128 to 1024.}
\label{fig:mpsi_vs_beta}
\end{figure}

In order to localize the critical coupling $\beta_{\rm c}^{(2)}$, we looked
for the crossing point at higher $\beta$ of the Binder cumulant
$B_4^{(M_R)}$ defined in Eq.~(\ref{binder_MR}) and repeated the 
analysis based on the optimal overlap of data points when they are
plotted against $(\beta-\beta_{\rm c})(\ln L)^{1/\nu}$, with 
$\nu$ fixed at 1/2. The same procedure was carried on using also
the observable $m_\psi$, which is itself an RG-invariant
quantity and shares therefore the same properties of a Binder cumulant
(see Fig.~\ref{fig:mpsi_vs_beta} for the behavior of $m_\psi$ versus $\beta$
on various lattices, which shows two crossing points, the one at higher
$\beta$ corresponding to the transition from the massless to the ordered
phase). This analysis led to the result $\beta^{(2)}_{\rm c}=1.1048(10)$, 
which agrees with the infinite volume extrapolation of the corresponding 
pseudocritical couplings. 

We can now determine the ratios of critical indices $\beta/\nu$ and 
$\gamma/\nu$ as we did in the previous Section. It should be noted, however,
that the population $S_L$ and its susceptibility are not suitable observables
for this purpose, being defined in a non-local manner and, thus, not
directly related to the two-point correlator. We use, instead, the rotated
magnetization $M_R$ and its susceptibility $\chi^{(M_R)}_L$ and compare their
values at the infinite volume critical coupling $\beta^{(2)}_{\rm c}$ (see
the last two columns of Table~\ref{beta2_pc}) with the scaling 
laws~(\ref{magn_fss}) and~(\ref{chiM_fss}), respectively.
Results for $\beta/\nu$ and $\gamma/\nu$ are summarized in 
Tables~~\ref{chiMR_second} and \ref{MR_second}.

All the values for $\eta^{(2)}=2-\gamma/\nu$ given in Table~\ref{chiMR_second}
are in agreement with the prediction $4/N^2$, which gives 0.16 for $N=5$. 
The hyperscaling relation $\gamma/\nu+2\beta/\nu=d$ is always satisfied,
within the statistical error.

The determination of the magnetic critical index based on the effective 
$\eta$ index defined in~(\ref{eff_eta_def}) is plagued, in this region
of values of $\beta$, by a sizeable dependence on the choice of the 
arbitrary parameter $R_0$. The shape of the curves for $\eta_{\rm eff}(R)$ 
and the way they depend on $R_0$ suggest that here logarithmic corrections
to the scaling could be at work. However, our data are not accurate enough to
include them reliably in our fits.

We conclude this Section by presenting several examples of {\it a posteriori}
check of consistency of our determinations for $\eta^{(1)}$ and 
$\eta^{(2)}$. The basic idea is to build plots in which we correlate
two RG-invariant quantities and to check that sequences of data points,
corresponding to different values of $\beta$, fall on a universal curve,
irrespective of the lattice size $L$~\cite{Challa:1986sk,Loison99}.

\begin{figure}[tb]
\vspace{0.5cm}
\centering
\includegraphics[scale=0.32]{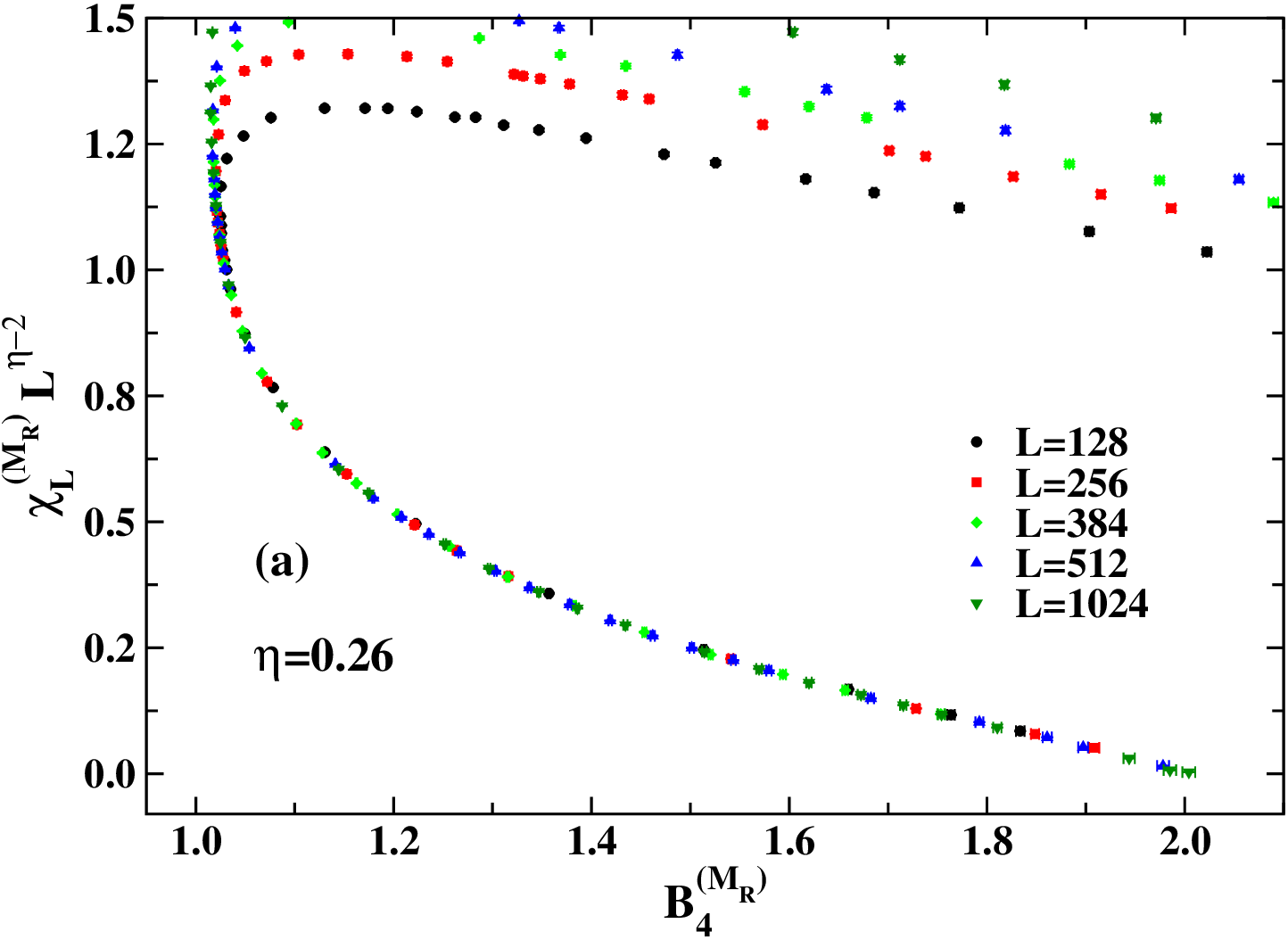}
\includegraphics[scale=0.32]{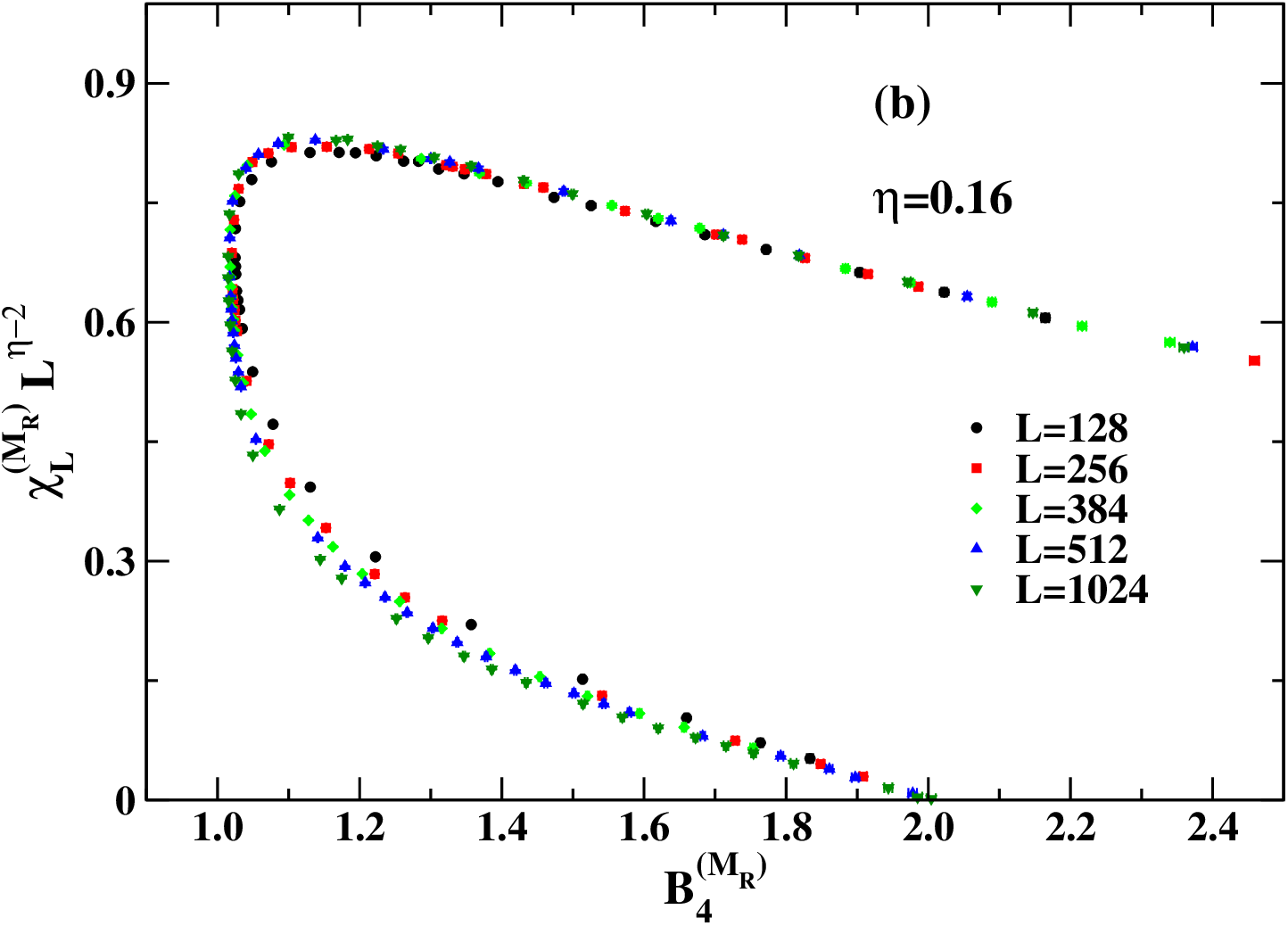}
\caption{(Color online) Correlation between $\chi_L^{(M_R)} L^{\eta-2}$ and the 
Binder cumulant $B_4^{(M_R)}$ for (a) $\eta=0.26$ and (b) $\eta=0.16$
on lattices with $L$ ranging from 128 to 1024. For $\eta=0.26$ 
(a) data from different lattices tend to fall on a universal curve in 
the lower branch, corresponding to $\beta$ values in the region of the first
transition. For $\eta=0.16$ (b) data from different lattices tend 
to fall on a universal curve in the upper branch, corresponding to $\beta$ 
values in the region of the second transition.}
\label{fig:chi_vs_b4}
\end{figure}

The first example is the plot of the rescaled susceptibility 
$\chi_L^{(M_R)} L^{\eta-2}$ against the Binder cumulant $B_4^{(M_R)}$.
One can see from Fig.~\ref{fig:chi_vs_b4}(a) that for 
$\eta=0.26\simeq\eta^{(1)}$ data points from different lattices fall on the
same curve in the lower branch, corresponding to $\beta$ values in the 
region of the first transition; for $\eta=0.16\simeq\eta^{(2)}$, on the
contrary, data points from different lattices fall on the same curve in the 
upper branch, corresponding to $\beta$ values in the region of the second
transition (see Fig.~\ref{fig:chi_vs_b4}(b)).

\begin{figure}[tb]
\vspace{0.5cm}
\centering
\includegraphics[scale=0.4]{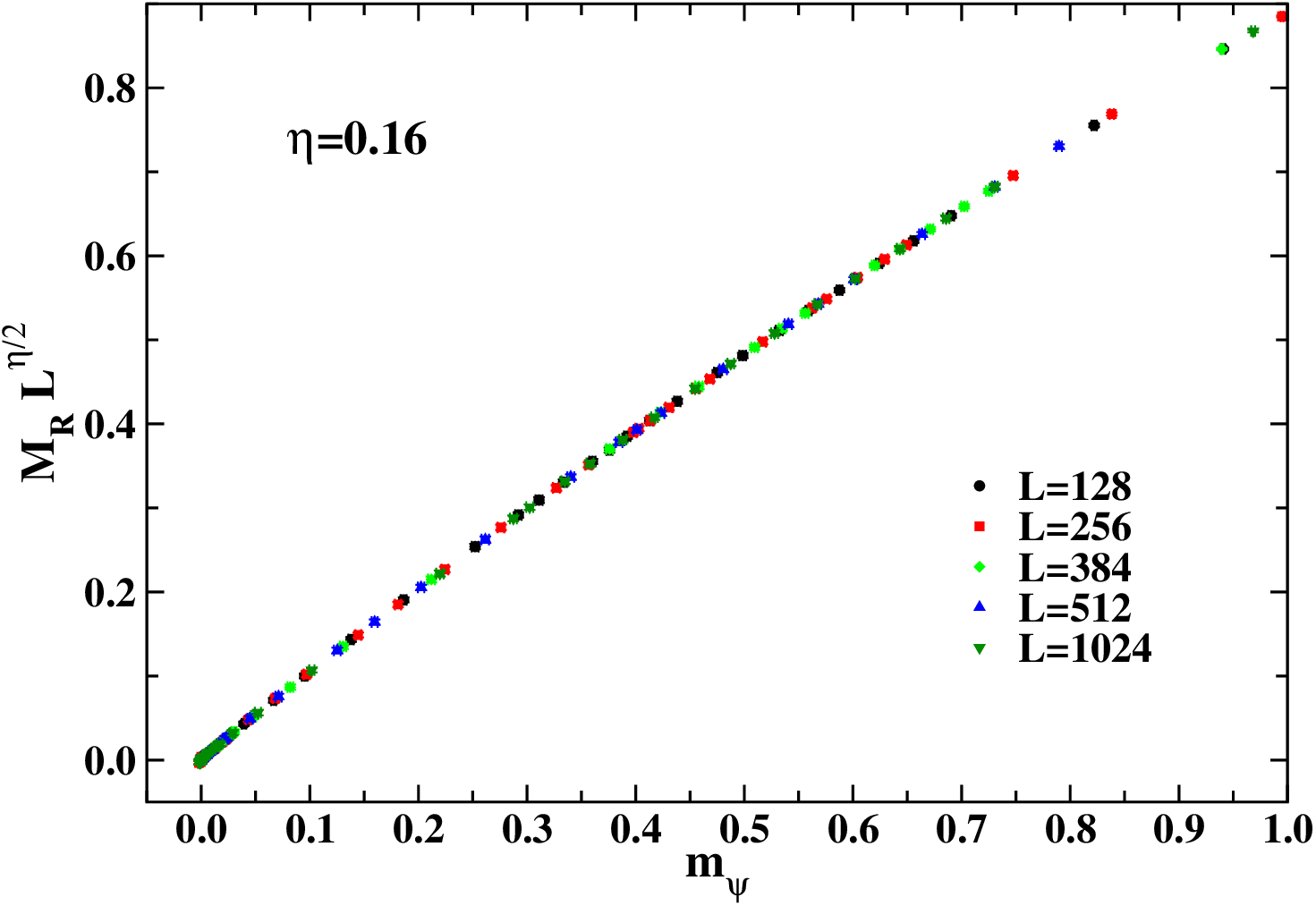}
\caption{(Color online) Correlation between $M_R L^{\eta/2}$ and $m_\psi$
for $\eta=0.16$ on lattices with $L$ ranging from 128 to 1024.}
\label{fig:MR_vs_mpsi}
\end{figure}

Another example is provided by the plot of the rescaled magnetization
$M_R L^{\eta/2}$ against $m_\psi$. 
For $\eta=0.16\simeq\eta^{(2)}$, again data points from different lattices 
fall on the same curve (see Fig.~\ref{fig:MR_vs_mpsi}).

\section{Discussion and conclusions}
\label{conclusions}

In this paper we have presented a wealth of numerical data aimed at 
shedding light on the phase structure of the $2D$ $Z(5)$ vector model. 
By means of a Monte Carlo cluster updating algorithm, designed to work
for $Z(N)$ models with odd $N$, we have outlined a scenario compatible
with the existence of three phases: disorder (small $\beta$), massless or
BKT (intermediate $\beta$), order (large $\beta$). We have determined
\begin{itemize}

\item the critical points $\beta^{(1)}_{\rm c}$ and $\beta^{(2)}_{\rm c}$
in the infinite volume limit, by means of the FSS of suitable Binder
cumulants;

\item the critical indices $\beta/\nu$ and $\gamma/\nu$ at the two 
critical points, by means of the FSS of suitable definitions of the
magnetization and of its susceptibility.
\end{itemize}

The determination of $\beta^{(2)}_{\rm c}$ has been cross-checked
with the infinite volume extrapolation of the (volume dependent) 
pseudocritical couplings of the second transition, assuming essential
scaling. We have found the following values of the critical couplings:
$\beta^{(1)}_{\rm c}=1.0510(10)$ and $\beta^{(2)}_{\rm c}=1.1048(10)$. 
As mentioned in the Introduction, values of the critical points are related by 
the duality transformations. In the Appendix we check, via the duality, 
the accuracy of our predictions and show that our determination of 
the critical couplings is in rather good agreement with it.

The determination of the index $\eta^{(1)}=2-\gamma/\nu$ at the 
first transition has been cross-checked with the effective $\eta$ index
method. The values of the index $\gamma/\nu$ at both critical points
agree well with theoretical predictions obtained for the Villain 
formulation~\cite{elitzur} thus supporting the conjecture that both standard 
and Villain formulations are in the same universality class. 
The behavior of the complex magnetization as well as the two-point 
correlation function strongly indicate that the intermediate phase 
is a massless phase whose symmetry is $U(1)$. Moreover, using 
spin-wave--vortex approximation and conventional perturbation theory 
one can calculate the two-point correlation function analytically 
and extract the perturbative beta-function in the intermediate phase. 
It turns out that the beta-function vanishes in this phase - a property 
which further supports the presence of the BKT transition and massless phase.

\begin{figure}[tb]
\centering
\includegraphics[scale=0.4]{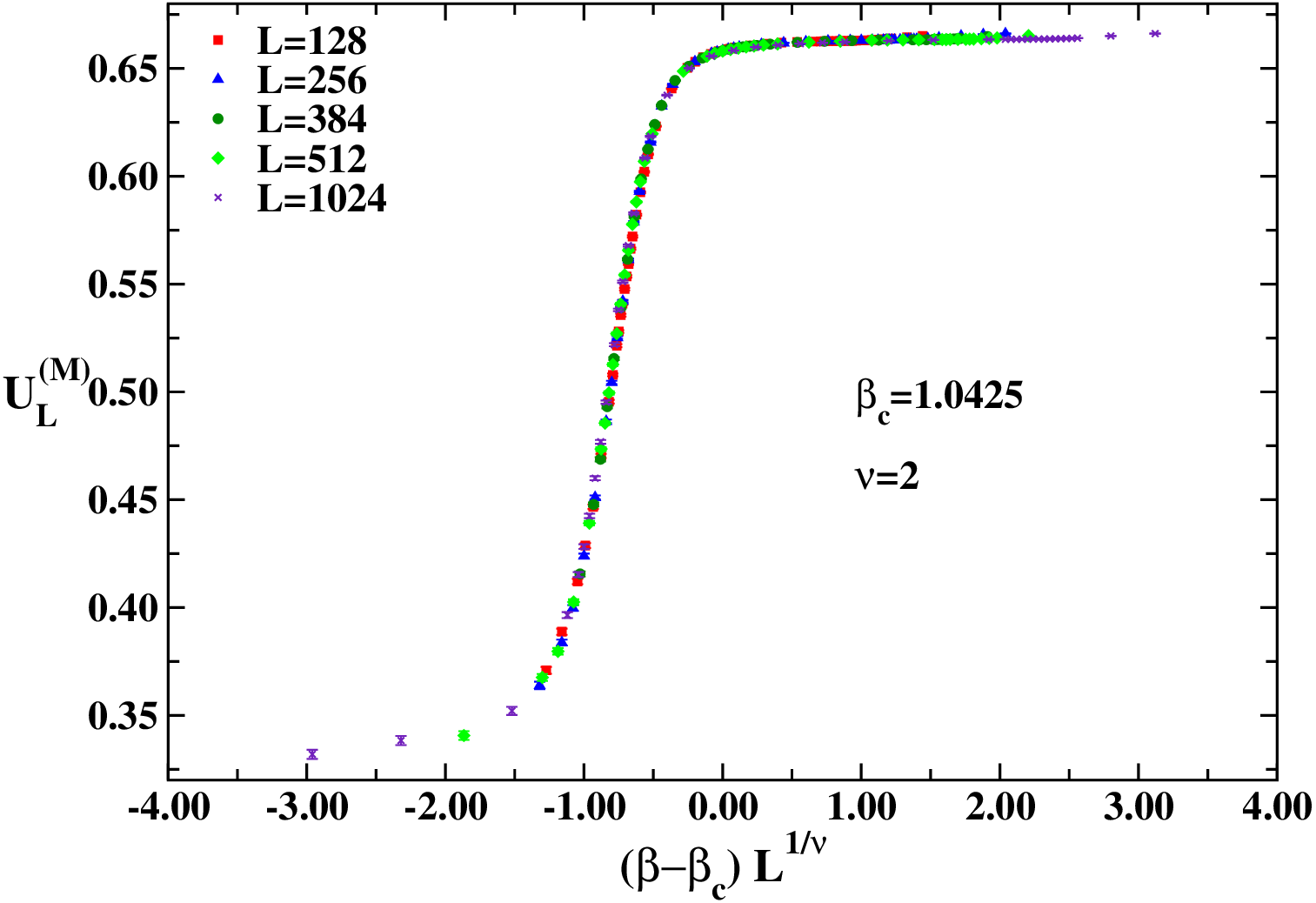}
\caption[]{(Color online) Reduced 4-th order Binder cumulant $U^{(M)}_L$ versus 
$(\beta-\beta_{\rm c})L^{1/\nu}$, for $\beta_{\rm c}=1.0425$ and
$\nu=2$ on lattices with $L$ ranging from 128 to 1024.}
\label{fig:binder_rescaled_II}
\end{figure}

For completeness, we have tested a scenario in which both transitions
are second order. In this case, the infinite volume extrapolation of the 
pseudocritical couplings should obey
\[
\beta^{(1,2)}_{\rm pc}=\beta^{(1,2)}_{\rm c}+\frac{A}{L^{\frac{1}{\nu}}}\quad ,
\]
whereas the overlap method of the Binder cumulants should work when data
are plotted against $(\beta-\beta_c) L^{\frac{1}{\nu}}$. Under these
conditions, we found
\begin{eqnarray*}
\beta^{(1)}_{\rm c}=1.0425(25) \;, & \;\;\; 1/\nu^{(1)}=0.50(5) \;,\\
\beta^{(2)}_{\rm c}=1.1075(25) \;, & \;\;\; 1/\nu^{(2)}=0.45(5) \;.
\end{eqnarray*}
In Fig.~\ref{fig:binder_rescaled_II} we show the overlap of the curves 
for the Binder cumulant $U_L^{(M)}$ obtained on various lattices when
plotted versus $(\beta-\beta_{\rm c})L^{1/\nu}$, with $1/\nu$ fixed at 0.5
and $\beta_{\rm c}=1.0425$. The quality of the overlap seems to be overall
better than in the BKT scenario of Fig.~\ref{fig:binder_rescaled}, but a 
closer inspection shows that it is worse in the region near the critical 
point.

\begin{figure}[tb]
\vspace{0.5cm}
\centering
\includegraphics[scale=0.4]{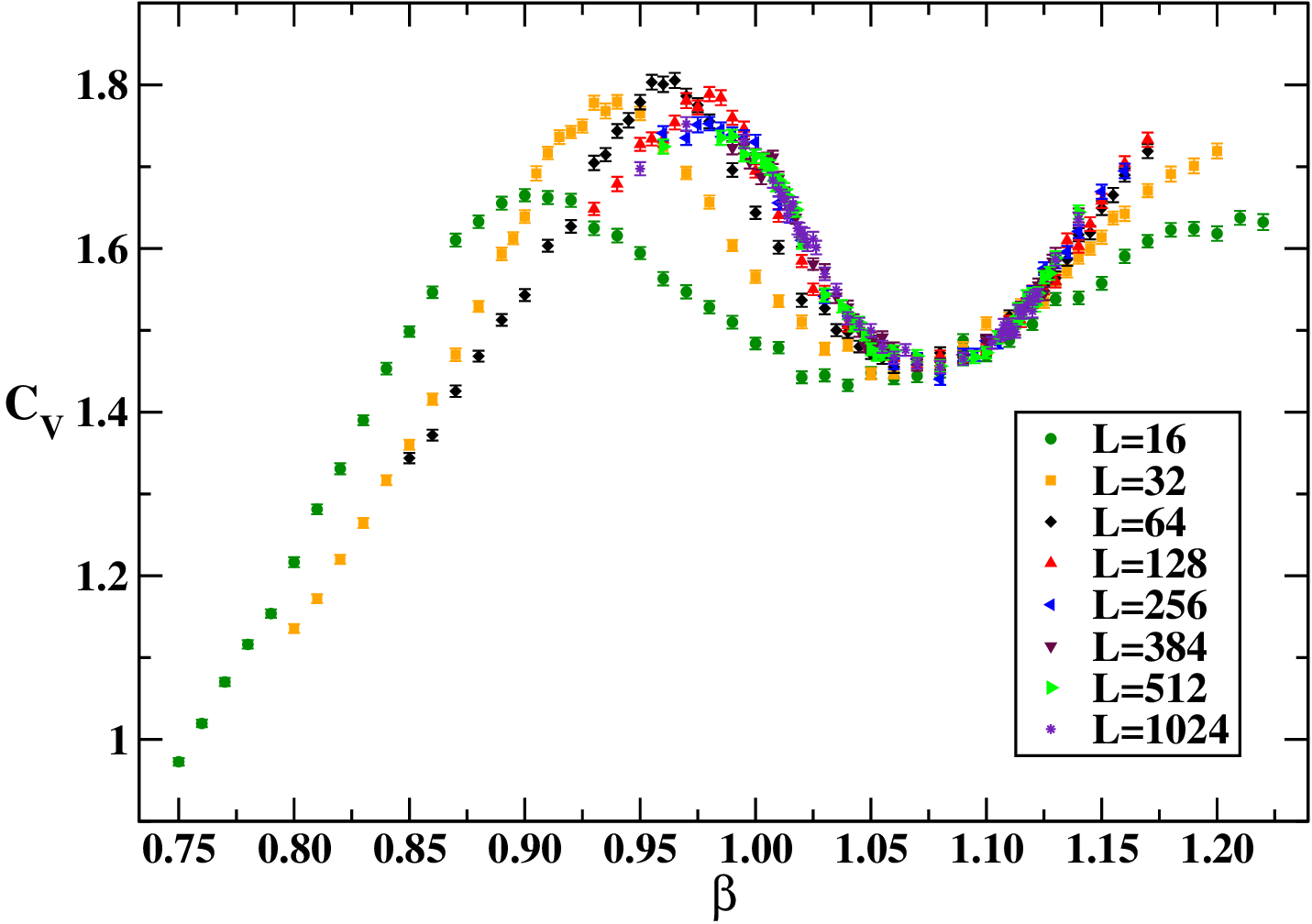}
\caption{(Color online) Specific heat $C_V$ measured on lattices with $L$ ranging from 
16 to 1024.}
\label{fig:C_V}
\end{figure}

Second order transition at $\beta^{(1)}_{\rm c}$, however, should be excluded 
by the numerical evidence that the intermediate phase is massless. In 
particular, two-point correlators tend to vanish at large distances for large 
volumes, whereas one should expect spontaneous symmetry breaking and 
non-vanishing values of long distance correlations in the case of second order 
phase transition. 
The exclusion of the second order transition at $\beta^{(2)}_{\rm c}$  
seems to be a more subtle problem, at least on the numerical side. 
One should probably simulate the system on larger lattices to reliably 
distinguish the BKT scenario from the second order one, if we deal with the 
quantities studied so far. However, one could consider more traditional 
observables to determine the order of the phase transition. 
The BKT phase transition is of 
infinite order. In particular, it is expected that the singular part of the 
free energy of the $XY$ model behaves like $F_s\sim\xi^{-2}$. Hence, all 
derivatives of the free energy are analytic functions of the temperature. 
In turn, the second derivative of the free energy shows a finite jump if 
the system undergoes a second order phase transition. We have decided, 
therefore, to compute the specific heat $C_V$ of the $Z(5)$ model. 
Figure~\ref{fig:C_V} shows the result of simulations for various lattice 
sizes. As suggested by this plot, the specific heat shows neither a divergence 
nor a finite jump. We interpret this behavior as further evidence in favor 
of the infinite order phase transition. On the theoretical side, the second 
order transition seems incompatible with the analysis of the dual 
transformations~\cite{Cardy,Domany}. 

As discussed in the Introduction, in a recent work~\cite{BM10} it has been 
claimed that the phase transition at 
$\beta^{(1)}_{\rm c}$  is not a standard BKT phase transition. The main 
tool in the analysis of Ref.~\cite{BM10} was the helicity modulus $\Upsilon$,
which was not considered in this work. The key observation in Ref.~\cite{BM10}
was that the helicity modulus does not jump to zero across the phase 
transition. 
This property prompted the authors of  Ref.~\cite{BM10} to conclude that 
the phase transition at $\beta^{(1)}_{\rm c}$ is a weaker cousin of the 
standard BKT transition. Our data indicate that there seem to be no influence 
of such behavior of the helicity modulus on other characteristic features 
of the phase transition. Most important on our opinion is the fact that 
both standard and Villain formulation are still in the same universality 
class since they show equal critical indices. Moreover, one could consider 
the critical index which governs the behavior of the helicity 
modulus~\cite{Fisher} 
$$
\Upsilon \ \sim \ \left ( \frac{T-T_c}{T_c} \right)^{\upsilon} \ , \;\;\;\;\; 
\upsilon = 2 \beta - \eta \nu \ .
$$
Our data for the indices $\beta$ and $\eta^{(1)}$ are compatible with a 
vanishing value 
of $\upsilon$. Thus, $\Upsilon=const$ right at the critical point. Obviously, 
$\upsilon=0$ also for the Villain model. The non-vanishing value of $\Upsilon$ 
at $\beta<\beta^{(1)}_{\rm c}$ seems to characterize rather the 
high-temperature phase than the massless BKT phase. Indeed, a physical 
interpretation given in Ref.~\cite{BM10} refers to lack of free vortices in 
the high-temperature phase of the $Z(5)$ models. It would then be interesting 
and important to study the dynamics of the vortex--anti-vortex pairs in both 
formulations. This dynamics can indeed be different. This task, however is 
beyond the scope of the present paper.  

\section*{Appendix} 

Consider general $Z(5)$ vector Potts model with Boltzmann weight given by 
\begin{equation*}
Q(s)\ =\ 1 + 2 x(\beta) \ \cos\frac{2\pi}{5} s + 2 y(\beta) 
\ \cos\frac{4 \pi}{5} s \ .
\label{Qpgen1}
\end{equation*}
Duality transformations read~\cite{Wuz5}
\begin{eqnarray*} 
x_d(\beta) & = & (1+2 x(\beta)\cos(2 \pi/5) + 2 y(\beta)\cos(4 \pi/5))/(1+2 x(\beta) + 2 y(\beta)) \ ,  \\
y_d(\beta) & =& (1+2 x(\beta)\cos(4 \pi/5) + 2 y(\beta)\cos(2 \pi/5))/(1+2 x(\beta) + 2 y(\beta)) \ .
\label{dualtr}
\end{eqnarray*}
The initial couplings $x(\beta)$ and $y(\beta)$ can be calculated from the 
Fourier transform of the original Boltzmann weight and in our case they are
given by
\begin{eqnarray} 
\label{xb}
x(\beta) \ &=& \ (-1-\sqrt{5}+(-1+\sqrt{5})e^{\frac{\sqrt{5}\beta}{2}}+ 
2 e^{\frac{1}{4}(5+\sqrt{5})\beta}) /(4+4 e^{\frac{\sqrt{5}\beta}{2}} + 
2 e^{\frac{1}{4}(5+\sqrt{5})\beta})  \ , \\
y(\beta) \ &=& \ (-1+\sqrt{5}-(1+\sqrt{5})e^{\frac{\sqrt{5}\beta}{2}}+ 
2 e^{\frac{1}{4}(5+\sqrt{5})\beta}) /(4+4 e^{\frac{\sqrt{5}\beta}{2}} + 
2 e^{\frac{1}{4}(5+\sqrt{5})\beta})  \ .
\label{yb}
\end{eqnarray}

It follows that 
\begin{eqnarray} 
\label{xd}
x_d(\beta) \ &=& \  e^{\frac{1}{4}(-5+\sqrt{5})\beta} \ , \\
y_d(\beta) \ &=& \  e^{- \frac{1}{4}(5+\sqrt{5})\beta} \ .
\label{yd}
\end{eqnarray}
Now, consider original and dual partition functions,
\begin{equation*}
 Z(x(\beta),y(\beta)) \ = \ C(\beta) \ Z(x_d(\beta),y_d(\beta)) \ .
\label{PF}
\end{equation*}
They have the same form and differ only by a smooth function $C(\beta)$.
Suppose the original partition function is critical at 
$\beta_c^{(1)}$ and $\beta_c^{(2)}$. These values correspond to 
$(x^{(1)}, y^{(1)})$ and $(x^{(2)}, y^{(2)})$. Let $(x_d^{(1)}, y_d^{(1)})$ 
and $(x_d^{(2)}, y_d^{(2)})$ be the values of dual couplings at critical 
points. The interaction in original and dual partition functions is the same.
Therefore, numerical values of the critical points in terms of $(x,y)$ and 
$(x_d,y_d)$ should be the same. However, we know from the solution of the 
self-dual equation (see \cite{Wuz5}) that the self-dual point is not a 
critical point for the vector Potts model. This leaves the only one
possibility, that at the critical points one must have 
\[
x^{(1)} = x_d^{(2)}\;, \;\;\;\;\; y^{(1)} = y_d^{(2)}
\]
and
\[
x^{(2)} = x_d^{(1)}\;, \;\;\;\;\: y^{(2)} = y_d^{(1)} \;.
\]
Results for critical points reported in the text are $\beta_c^{(1)} = 1.051$ 
and $\beta_c^{(2)} = 1.1048$. So, we easily find from~(\ref{xb})-(\ref{yd})
\begin{eqnarray*}
x(1.051) = 0.466532\;, && x_d(1.1048) = 0.46608 \\
y(1.051) = 0.136626\;, && y_d(1.1048) = 0.135523
\end{eqnarray*}
and 
\begin{eqnarray*}
x(1.1048) = 0.485097\;, && x_d(1.051) = 0.483733 \\
y(1.1048) = 0.149612\;, && y_d(1.051) = 0.149378 \;.
\end{eqnarray*}

\section*{Acknowledgments}

O.B. thanks for warm hospitality the Dipartimento di Fisica dell'Universit\`a 
della Ca\-la\-bria and the INFN Gruppo Collegato di Cosenza during the work on 
this paper. G.C. and A.P. thank the BITP, Kiev for friendly hospitality
during the last stages of this work. Numerical simulations were performed on 
the linux PC farm ``Majorana'' of the INFN-Cosenza and on the GRID cluster at 
the BITP, Kiev.

\end{document}